\documentclass[twocolumn]{aastex62}

\graphicspath{{./}{figures/}}

\newcommand{\angstrom}{{\rm \AA}}

\newcommand{\hbeta}{H{$\beta$}}
\newcommand{\halpha}{H{$\alpha$}}

\newcommand{\HII}{H{\sevenrm\,II}}

\newcommand{\OIII}{[O{\sevenrm\,III}]}

\newcommand{\OIIIb}{[O{\sevenrm\,III}]\,$\lambda$5007}

 \font\sevenrm=cmr7 scaled 1000

\newcommand{\chandra}{{\it Chandra}}

\shorttitle{SDSS AGN Pairs. III. X-ray Confirmation}
\shortauthors{Hou et al.}

\usepackage{graphicx}

\begin{document}

\title{Active Galactic Nucleus Pairs from the Sloan Digital Sky Survey. III. \chandra\ X-ray Observations Unveil Obscured Double Nuclei}

\email{houmc@smail.nju.edu.cn, xinliuxl@illinois.edu}

\author{Meicun Hou}
\affiliation{School of Astronomy and Space Science, Nanjing University, Nanjing 210046, China}
\affiliation{Key Laboratory of Modern Astronomy and Astrophysics (Nanjing University), Ministry of Education, Nanjing 210046, China}
\affiliation{Department of Astronomy, University of Illinois at Urbana-Champaign, Urbana, IL 61801, USA}

\author{Xin Liu}
\affiliation{Department of Astronomy, University of Illinois at Urbana-Champaign, Urbana, IL 61801, USA}
\affiliation{National Center for Supercomputing Applications, University of Illinois at Urbana-Champaign, 605 East Springfield Avenue, Champaign, IL 61820, USA}

\author{Hengxiao Guo}
\affiliation{Department of Astronomy, University of Illinois at Urbana-Champaign, Urbana, IL 61801, USA}
\affiliation{National Center for Supercomputing Applications, University of Illinois at Urbana-Champaign, 605 East Springfield Avenue, Champaign, IL 61820, USA}

\author{Zhiyuan Li}
\affiliation{School of Astronomy and Space Science, Nanjing University, Nanjing 210046, China}
\affiliation{Key Laboratory of Modern Astronomy and Astrophysics (Nanjing University), Ministry of Education, Nanjing 210046, China}

\author{Yue Shen}
\altaffiliation{Alfred P. Sloan Research Fellow}
\affiliation{Department of Astronomy, University of Illinois at Urbana-Champaign, Urbana, IL 61801, USA}
\affiliation{National Center for Supercomputing Applications, University of Illinois at Urbana-Champaign, 605 East Springfield Avenue, Champaign, IL 61820, USA}

\author{Paul J. Green}
\affiliation{Harvard-Smithsonian Center for Astrophysics, 60 Garden Street, Cambridge, MA 02138, USA}

\begin{abstract}
We present \chandra\ ACIS-S X-ray imaging spectroscopy for five dual active galactic nucleus (AGN) candidates. Our targets were drawn from a sample of 1286 \OIII -selected AGN pairs systematically selected from the Sloan Digital Sky Survey Seventh Data Release. Each of the targets contains two nuclei separated by $\sim$3--9 kpc in projection, both of which are optically classified as Type 2 (obscured) AGNs based on diagnostic ratios of the narrow emission lines. Combined with independent, empirical star formation rate estimates based on the host-galaxy stellar continua, the new \chandra\ X-ray observations allow us to evaluate the dual-AGN hypothesis for each merging system. We confirm two (SDSS J0907+5203 and SDSS J1544+0446) of the five targets as bona-fide dual AGNs. For the other three targets, the existing data are consistent with the dual-AGN scenario, but we cannot rule out the possibility of stellar/shock heating and/or one AGN ionizing both gaseous components in the merger. The average X-ray-to-\OIII\ luminosity ratio in our targets seems to be systematically smaller than that observed in single AGNs but is higher than that seen in dual AGNs selected from AGNs with double-peaked narrow emission lines. We suggest that the systematically smaller X-ray-to-\OIII\ luminosity ratio observed in dual AGNs than in single AGNs is due to a high nuclear gas column likely from strong merger-induced inflows. Unlike double-peaked-\OIII -selected dual AGNs, the new sample selected from resolved galaxy pairs are not subject to the orientation bias caused by the double-peak line-of-sight velocity splitting selection, which also contributes to lowering the X-ray-to-\OIII\ luminosity ratio. 

\end{abstract}


\section{Introduction} \label{sec:intro}

The observed growth of structures suggests that mergers of galaxies, and by extension, their central supermassive black holes \citep[SMBHs;][]{kormendy95,ff05}, should be common throughout most of cosmic history. The final coalescence of merging SMBHs would produce low-frequency gravitational waves \citep[GWs; e.g.,][]{colpi09}, providing a ``standard siren'' for cosmology \citep{hughes09} and a direct testbed for strong-field General Relativity \citep{Centrella2010}. Unlike LIGO's stellar mass black holes \citep{LIGO2016} whose detection is still limited to the relatively nearby universe, merging SMBHs would be detectable close to the edge of the observable universe \citep{Cornish2018}. They are being hunted currently by pulsar timing arrays for the more massive, low-redshift population \citep[e.g.,][]{Arzoumanian2018a}, and in the future by space-borne experiments for the less massive, high-redshift population \citep[e.g.,][]{Amaro-Seoane2017}. 

While the GWs from merging SMBHs are yet to be detected, it is useful to study their progenitors -- pairs of SMBHs selected by their electromagnetic signatures from accreting materials from the surroundings. These so-called ``dual active galactic nuclei (AGNs)'' are AGN pairs in merging galaxies with typical separations of a few kiloparsecs \citep{gerke07,comerford08,xu09}. Dual AGNs and AGN pairs in general provide an exciting prospect for understanding massive black hole growth and their merger rates in galaxies in the era of multi-messenger astronomy \citep[e.g.,][]{Bhowmick2019}. The first concrete evidence for dual AGNs was a serendipitous discovery by \chandra\ in NGC 6240 \citep{komossa03}, a $z=0.02$ merging ultra-luminous infrared galaxy (ULIRG), which contains two X-ray nuclei separated by $\sim 1\arcsec$ ($\sim 0.7$\,kpc) in projection. Until recently only a handful of other secure cases were known in the X-rays (e.g., 3C 75 \citep{hudson06}, Mrk 463 \citep{bianchi08,Treister2018}, Mrk 266 \citep{brassington07,mazzarella11}, and Mrk 739 \citep{Koss2011}; but see also Arp 299 \citep{ballo04} for a candidate), all of which were confirmed by \chandra.

The past decade has seen significant progress in finding concrete evidence for dual AGNs at $z<0.5$ thanks to systematic searches using large surveys combined with dedicated follow ups, in particular in the X-rays, mid-IR, and radio \citep[e.g.,][]{green10,fabbiano11,Liu2013,Comerford2015,Kocevski2015,Fu2015a,Gatti2016,Koss2016,Satyapal2017,Ellison2017,Fu2018,DeRosa2018,Liu2018,Vignali2018,Pfeifle2019}. \citet{Liu2011a} identified a sample of 1286 \OIII -selected, spatially resolved AGN pairs from the Sloan Digital Sky Survey (SDSS) Seventh Data Release \citep[DR7;][]{SDSSDR7}. Among the 1286 pairs, 94 have projected nuclear separation $r_{\rm p} <10$ kpc, making them the largest sample of \OIII -selected dual AGN candidates. Each of these galaxies contains two nuclei separated by a few arcsec, both of which have individual SDSS spectra. Due to the finite size of SDSS fibers (i.e., fiber collisions), two objects separated by $<55$'' cannot both be spectroscopically observed unless being on overlapping plates. We have corrected for this spectroscopic incompleteness when calculating the frequency of dual AGNs \citep{Liu2011a}. Based on the SDSS spectra, both nuclei in the candidates are optically classified as Type 2 (i.e., obscured) AGNs according to diagnostic ionization ratios of narrow emission lines \citep{bpt,veilleux87,kewley01}. However, optical diagnostic line ratios only represent indirect evidence for dual AGNs; they cannot conclusively rule out alternative scenarios for the nature of the ionizing sources such as stellar/shock heating, and/or a pair of merging galaxy components ionized by a single active nucleus.

It is generally thought that the X-ray band can provide more direct evidence for nuclear activity. In particular, hard X-rays (defined here as 2-10 keV) are transparent to column densities of $N_{\rm H} \lesssim 10^{24}\ {\rm cm^{-2}}$, while Compton-thick AGNs can be revealed at energies $\gtrsim$10 keV \citep{LaMassa2011,Lansbury2015,Nardini2017}.

\begin{figure}
 \centering
    \includegraphics[width=0.45\textwidth, angle=180]{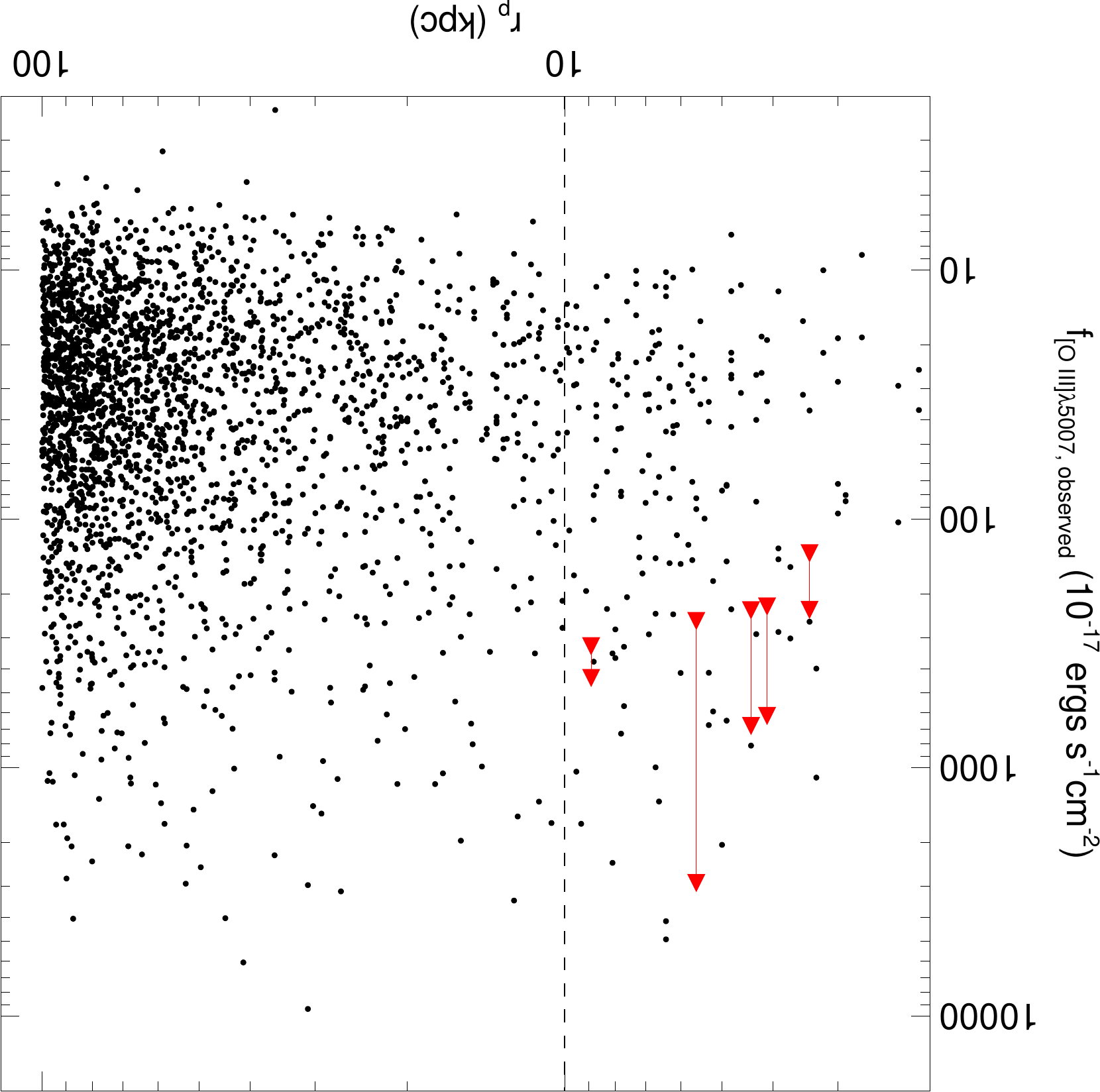}
    \caption{\OIII~flux vs. projected separation $r_{\rm p}$ for our five target pairs and parent sample of 1286 AGN pairs \citep{Liu2011a} in red triangles and black dots, respectively. Our five targets are selected by requiring $r_{\rm p} < 10$ kpc and both nuclei have sufficiently high \OIII~fluxes. The vertical dashed line marks the position of $r_{\rm p} = 10$ kpc.}
    \label{fig:sample}
\end{figure}

\begin{figure*}
 \centering
 \includegraphics[width=1.0\textwidth]{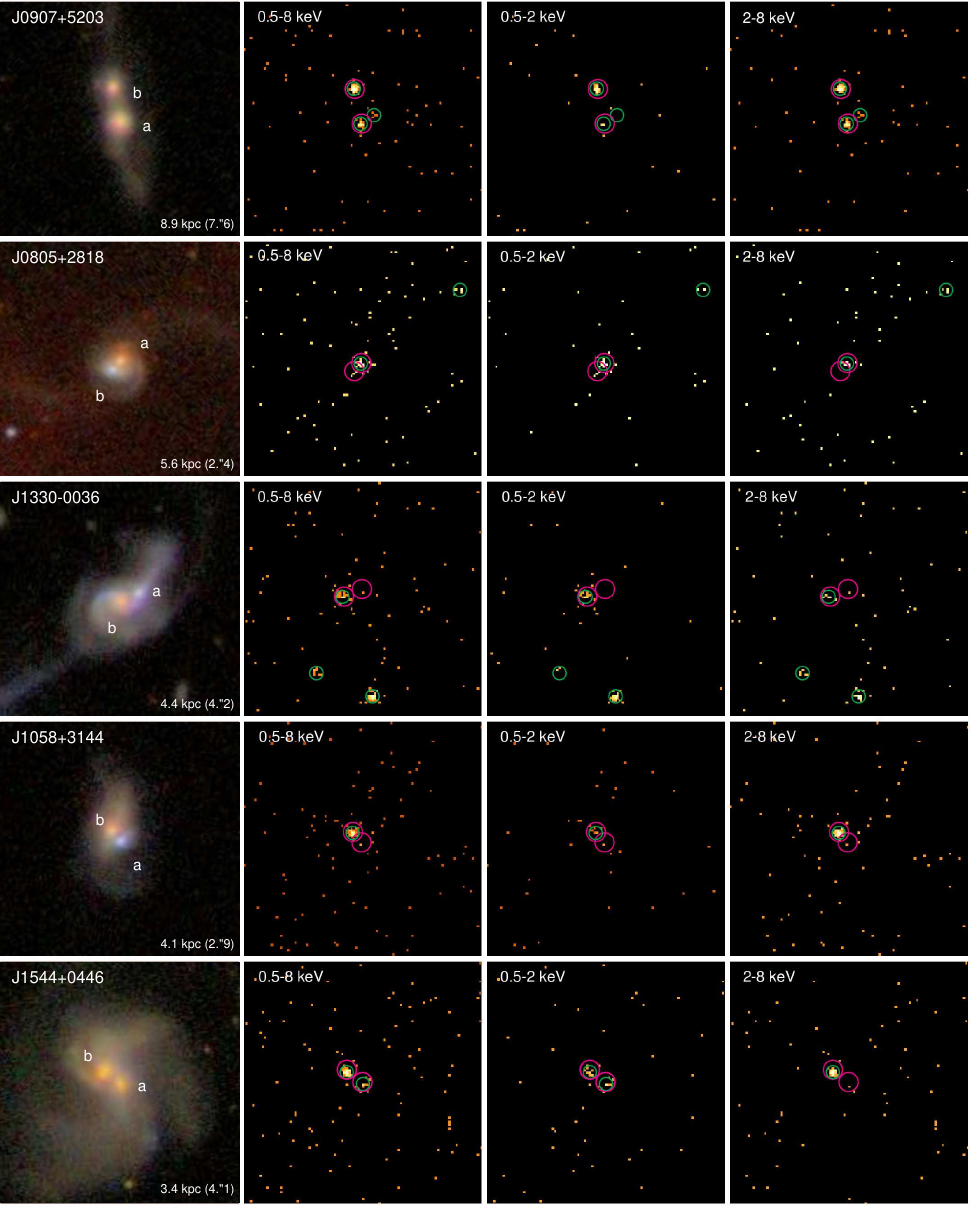}
    \caption{
    SDSS $gri$-color composite, \chandra\ 0.5--8 keV, 0.5--2 keV, and 2--8 keV images of the five \OIII -selected dual-AGN candidates. Each panel is $50''\times50''$. North is up and east is to the left. The targets are ordered with decreasing projected separation (as labeled, with angular distance in parenthesis) from top to bottom.
   Magenta circles denote positions of the optical nuclei whereas green circles represent the 90\% EER of the detected X-ray sources.
   Table 1 lists their basic properties.
    }\label{fig:sdssimg}
\end{figure*}

We here present a new \chandra\ ACIS-S X-ray imaging of five dual AGN candidates drawn from the parent sample of 94 closely separated, dual AGN candidates (Figure \ref{fig:sample}). Figure \ref{fig:sdssimg} shows their optical and X-ray images. \chandra~'s high image quality (FWHM $\lesssim 1''$) and its capability of moderate spectral resolution imaging spectroscopy in the X-rays make it ideal for assessing the dual-AGN hypothesis. Unlike previous \chandra~'s searches for dual AGNs in galaxy pairs hosting single AGNs \citep[e.g.,][]{teng12}, we target merging galaxies where {\it both} nuclei are \OIII -selected AGNs.


The paper is organized as follows. Section \ref{sec:target_selection} describes the target selection and their optical properties. Section \ref{sec:obs} presents the new \chandra\ ACIS-S X-ray observations, data reduction, and data analysis. Section \ref{sec:result} shows the results on the X-ray luminosities and spectral properties, the X-ray contribution from star formation, and the nature of the nuclear ionizing sources. Finally, we discuss the implications of our results in Section \ref{sec:discuss} and summarize the main conclusions in Section \ref{sec:sum}. Throughout this paper, we assume a concordance cosmology with $\Omega_m = 0.3$, $\Omega_{\Lambda} = 0.7$, and $H_{0}=70$ km s$^{-1}$ Mpc$^{-1}$, and use the AB magnitude system \citep{oke74}.

\section{Target Selection and Optical Properties}\label{sec:target_selection}

Our targets are drawn from a parent sample of 1286 spectroscopic AGN pairs with $r_{\rm p} < 100$ kpc and line-of-sight velocity offsets $<$ 600 km s$^{-1}$ selected from a heterogenous sample of 138,070 optical AGNs from the SDSS DR7. The optical AGN sample consists of 129,277 Type 2 (i.e., narrow-line) AGNs, 5,564 Type 1 (i.e., broad-line) AGNs, 3,117 Type 1 quasars, and 112 Type 2 quasars \citep{Liu2011a}. The parent AGN pair sample is dominated by Type 2 AGNs whose optical narrow emission-line ratios are characteristic of Seyferts, LINERs, and/or AGN-\HII\ composites. 

\begin{figure*}
 \centering
    \includegraphics[width=0.95\textwidth]{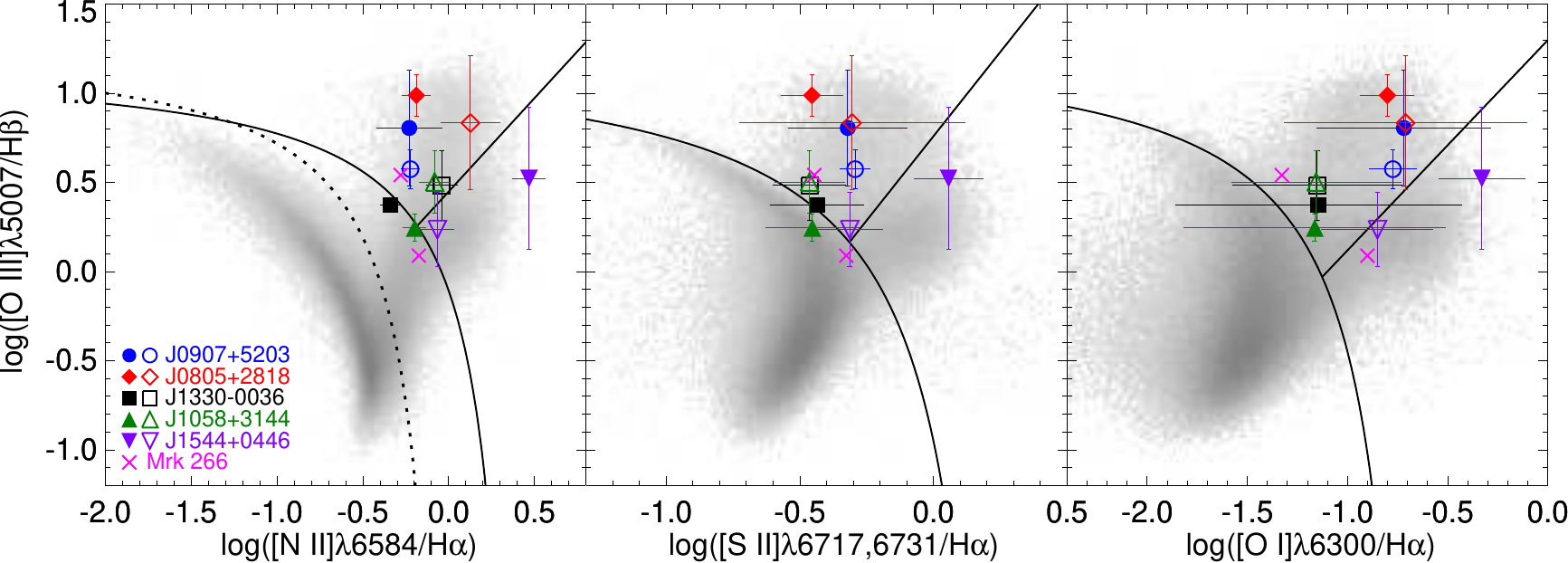}
    \caption{
     Optical diagnostic emission-line ratios \citep{bpt,veilleux87} of the 10 nuclei in our five dual AGN candidates (colored solid symbols) and of the two nuclei in Mrk 266 (red crosses).
     Gray scales indicate number densities of SDSS DR4 emission-line galaxies \citep{kauffmann03}.
     The dashed curve denotes the empirical separation between H {\tiny II} regions and AGNs \citep{kauffmann03}, the solid curve displays the theoretical ``starburst limit'' \citep{kewley01}, and the solid line represents the empirical division between Seyferts and LINERs \citep{kewley06}.
     Pure star-forming (``SF'') galaxies lie below the dashed curve, AGN-dominated objects (Seyferts above and LINERs below the solid line) lie above the solid curve, and AGN-H {\tiny II} composites lie in between.
    }\label{fig:bpt}
\end{figure*}

From the parent AGN pair sample, we select kpc-scale dual AGN candidates by first requiring $r_{\rm p} < 10$ kpc. We further select systems in which both nuclei have high enough \OIII\ fluxes to expect at least 50 hard (2-10 keV) X-ray counts in 15 ks for the weaker component (see details below). We additionally exclude systems with existing {\it Chandra} observations to avoid duplicate observations (e.g., Mrk 266). For systems with similar \OIII\ fluxes, targets which are classified as Seyferts based on the BPT diagram are prioritized over those that are classified as AGN-H {\tiny II} composites to maximize the probability of X-ray detections. Therefore, our final target sample consists of five dual AGN candidates.
The distribution of our sample is shown in Figure \ref{fig:sample}.
Tables \ref{table:obs} and \ref{table:emission_line} list their basic photometric and spectroscopic properties. The stellar velocity dispersion was measured by fitting the host-galaxy stellar continuum using the penalized Pixel-Fitting (pPXF) method \citep{Cappellari2004} (see the Appendix for details). The total stellar mass was given by the MPA-JHU DR7 catalog from fitting the photometry \citep{kauffmann03c,salim07}. Figure \ref{fig:bpt} shows the narrow emission-line ratios measured from the SDSS spectra subtracted for stellar continua using the pPXF fits. It illustrates that the nuclei in our targets are optically classified as Type 2 Seyferts, LINERs, or AGN-\HII\ composites. 


\begin{deluxetable*}{lcccccccccc}
\tablecaption{\OIII -Selected Dual AGN Candidates with New {\it Chandra} ACIS-S X-ray Observations. \label{table:obs}}
\tablehead{
\colhead{Full Name} & 
\colhead{Name} & 
\colhead{Redshift} & 
\colhead{$r_{\rm p}$} & 
\colhead{Plate} & 
\colhead{Fiber ID} & 
\colhead{MJD} &  
\colhead{$m_r$} & 
\colhead{$\sigma_{\ast}$} & 
\colhead{log$M_{{\rm BH}}$} & 
\colhead{log$M_{\ast}$} 
}
\colnumbers
\startdata
J090714.45+520343.4 & J0907+5203a & 0.0596 & 8.9& 553 & 208 & 51999 & 16.27 & 177 & 7.9 & 10.6  \\
J090714.61+520350.7 & J0907+5203b & 0.0602 &  8.9& 552 & 603 & 51992 & 16.95 & 132 & 7.4 & 10.3  \\
\hline
J080523.29+281815.8 & J0805+2818a & 0.1284 & 5.6& 929 & 570 & 52581 & 15.99 & 316 & 9.0 & 11.2 \\
J080523.40+281814.1 & J0805+2818b & 0.1286 & 5.6& 930 & 285 & 52618 & 18.18 & 169 & 7.8 & 9.8  \\
\hline
J133031.75$-$003611.9 & J1330$-$0036a & 0.0542 & 4.4& 298 & 264 & 51955 & 18.36 & 72 & 6.2 & 8.8 \\
J133032.00$-$003613.5 & J1330$-$0036b & 0.0542 & 4.4& 297 & 61 & 51959 & 15.17 & 153 & 7.6 & 10.7 \\
\hline
J105842.44+314457.6 & J1058+3144a & 0.0728 & 4.1& 1990 & 342 & 53472 & 17.27 & 101 & 6.9 & 10.0  \\
J105842.58+314459.8 & J1058+3144b & 0.0723 & 4.1& 2026 & 17 & 53711 & 15.79 & 159 & 7.7 & 10.9 \\
\hline
J154403.45+044607.5 & J1544+0446a & 0.0420 & 3.4& 2950 & 278 & 54559 & 17.82 & 153 & 7.6 & 9.8  \\
J154403.67+044610.1 & J1544+0446b & 0.0416 & 3.4& 1836 & 586 & 54567 & 14.21 & 201 & 8.1 & 11.1 \\
\enddata
\tablecomments{(1) SDSS names with J2000 coordinates given in the form of ``hhmmss.ss+ddmmss.s''; (2) Abbreviated Name, a and b for each dual AGN candidate are ordered with R.A.; (3) Spectroscopic redshift from the SDSS DR7; (4) Projected physical separation of dual AGN in each pair, in units of kpc; (5)-(7) SDSS spectroscopic plate number, fiber ID, and Modified Julian Date; (8) SDSS r-band model magnitude; (9) Stellar velocity dispersion provided in the MPA-JHU DR7 catalog, in units of km s$^{-1}$; (10) Black hole mass estimate inferred from $\sigma_{\ast}$ assuming the $M_{{\rm BH}}$-$\sigma_{\ast}$ relation of \cite{Gultekin2009}, in units of ${\rm M_\odot}$; (11) Total stellar mass from fits to the photometry provided in the MPA-JHU DR7 catalog \citep{salim07}, in units of ${\rm M_\odot}$. 
}
\end{deluxetable*}

\begin{deluxetable*}{lcccccccccc}
\tablecaption{Optical Emission-line Properties of the Five \OIII -Selected Dual AGN Candidates. \label{table:emission_line}}
\tabletypesize{\scriptsize}
\tablehead{
\colhead{Name} & 
\colhead{${{\rm H}\beta}$} & 
\colhead{${{\rm H}\alpha}$} & 
\colhead{${\rm [O~III]}$} & 
\colhead{${\rm [N~II]}$} & 
\colhead{${\rm [S~II]}$} &  
\colhead{${\rm [O~I]}$} & 
\colhead{$\frac{{\rm [O~III]}}{{\rm H}\beta}$} & 
\colhead{$\frac{{\rm [N~II]}}{{\rm H}\alpha}$} & 
\colhead{$\frac{{\rm [S~II]}}{{\rm H}\alpha}$} & 
\colhead{$\frac{{\rm [O~I]}}{{\rm H}\alpha}$} 
}
\colnumbers
\startdata
J0907+5203a & $   98.8\pm  70.9$ & $  361.6\pm  81.5$ & $  632.4\pm 141.8$ & $  213.0\pm  81.5$ & $  172.8\pm  80.2$ & $   69.1\pm  67.4$ & $    6.4\pm   4.8$ & $    0.6\pm   0.3$ & $    0.5\pm   0.2$ & $    0.2\pm   0.2$ \\
J0907+5203b & $  130.9\pm  29.9$ & $  581.7\pm  40.6$ & $  492.3\pm  49.1$ & $  348.1\pm  36.3$ & $  296.2\pm  33.2$ & $   98.0\pm  25.6$ & $    3.8\pm   0.9$ & $    0.6\pm   0.1$ & $    0.5\pm   0.1$ & $    0.2\pm   0.0$ \\
\hline
J0805+2818a & $  390.0\pm 102.2$ & $ 1578.6\pm 180.2$ & $ 3808.5\pm 266.3$ & $ 1023.7\pm 161.4$ & $  554.2\pm 135.7$ & $  249.4\pm  72.6$ & $    9.8\pm   2.6$ & $    0.6\pm   0.1$ & $    0.4\pm   0.1$ & $    0.2\pm   0.0$ \\
J0805+2818b & $   52.7\pm  43.0$ & $  150.4\pm  40.2$ & $  360.1\pm 103.3$ & $  201.1\pm  60.3$ & $   74.6\pm  70.1$ & $   29.3\pm  40.2$ & $    6.8\pm   5.9$ & $    1.3\pm   0.5$ & $    0.5\pm   0.5$ & $    0.2\pm   0.3$ \\
\hline
J1330$-$0036a & $  111.6\pm   8.2$ & $  398.8\pm  49.3$ & $  264.2\pm  12.2$ & $  182.0\pm  10.2$ & $  146.1\pm  56.9$ & $   28.5\pm  46.9$ & $    2.4\pm   0.2$ & $    0.5\pm   0.1$ & $    0.4\pm   0.1$ & $    0.1\pm   0.1$ \\
J1330$-$0036b & $  367.6\pm 144.1$ & $ 1568.2\pm 227.7$ & $ 1120.1\pm 251.0$ & $ 1426.4\pm 218.5$ & $  539.2\pm 151.3$ & $  110.7\pm 106.9$ & $    3.0\pm   1.4$ & $    0.9\pm   0.2$ & $    0.3\pm   0.1$ & $    0.1\pm   0.1$ \\
\hline
J1058+3144a & $  129.7\pm  15.3$ & $  377.3\pm  50.1$ & $  228.2\pm  30.0$ & $  239.2\pm  20.8$ & $  132.9\pm  51.0$ & $   25.9\pm  38.8$ & $    1.8\pm   0.3$ & $    0.6\pm   0.1$ & $    0.4\pm   0.1$ & $    0.1\pm   0.1$ \\
J1058+3144b & $  238.7\pm  83.8$ & $  921.0\pm 127.2$ & $  759.7\pm 145.9$ & $  762.2\pm 117.9$ & $  316.4\pm  85.6$ & $   64.2\pm  62.1$ & $    3.2\pm   1.3$ & $    0.8\pm   0.2$ & $    0.3\pm   0.1$ & $    0.1\pm   0.1$ \\
\hline
J1544+0446a & $   46.1\pm  38.5$ & $  209.1\pm  40.6$ & $  153.7\pm  57.7$ & $  613.8\pm  68.4$ & $  238.2\pm  54.4$ & $   97.9\pm  44.9$ & $    3.3\pm   3.1$ & $    2.9\pm   0.7$ & $    1.1\pm   0.3$ & $    0.5\pm   0.2$ \\
J1544+0446b & $  165.8\pm  62.6$ & $  749.4\pm 122.0$ & $  285.9\pm  85.7$ & $  643.3\pm 105.5$ & $  364.2\pm  86.3$ & $  105.7\pm  65.9$ & $    1.7\pm   0.8$ & $    0.9\pm   0.2$ & $    0.5\pm   0.1$ & $    0.1\pm   0.1$ \\
\enddata
\tablecomments{(2)-(7) Optical emission-line fluxes and 1$\sigma$ uncertainties in units of $10^{-17}{\rm~erg~s^{-1}~cm^{-2}}$; (8)-(11) Optical diagnostic emission-line ratios and 1$\sigma$ uncertainties estimated from error propagation. 
}
\end{deluxetable*}

\begin{deluxetable*}{lccccccccc}
\tablecaption{\chandra\ Observations of the Five \OIII -Selected Dual AGN Targets. \label{table:xray}}
\tabletypesize{\footnotesize}
\tablehead{
\colhead{Name} & 
\colhead{ObsID} & 
\colhead{Exp.} & 
\colhead{Counts} & 
\colhead{CR} & 
\colhead{CR1} &  
\colhead{CR2} & 
\colhead{HR} & 
\colhead{$\Gamma $} & 
\colhead{${\rm~N_{H}}$} 
}
\colnumbers
\startdata
J0907+5203a  & 14965 & 14.5& 33.9  & $ 2.76\pm 0.44$ & $ 0.33\pm 0.15$ & $ 2.43\pm 0.42$ & $0.76^{+0.12}_{-0.10}$   & - & -  \\
J0907+5203b* & 14965 & 14.5& 108.9 & $ 7.63\pm 0.73$ & $ 1.57\pm 0.32$ & $ 6.06\pm 0.66$ & $0.56^{+0.08}_{-0.07}$  & $1.43^{+0.84}_{-0.75}$ & $2.6^{+1.5}_{-1.1}$ \\
\hline
J0805+2818a  & 14964 & 14.0& 28.9  & $ 2.04\pm 0.38$ & $ 1.16\pm 0.28$ & $ 0.89\pm 0.26$ & $-0.17^{+0.18}_{-0.19}$ & - & -  \\
J0805+2818b  & 14964 & 14.0& $<$ 12.4 & $<$ 0.87&	$<$ 0.88&	$<$ 0.45& - & - & -   \\
\hline
J1330$-$0036a & 14967 & 14.9& $<$ 8.4 & $<$ 0.57&	$<$ 0.43&	$<$ 0.59& - & - & - \\
J1330$-$0036b & 14967 & 14.9& 15.9 & $ 1.05\pm 0.27$ & $ 0.57\pm 0.19$ & $ 0.48\pm 0.18$ & $-0.13^{+0.27}_{-0.28}$ & - & -  \\
\hline
J1058+3144a  & 14966 & 14.5& $<$ 12.1 & $<$ 0.81&	$<$ 0.72&	$<$ 0.57& - & - & - \\
J1058+3144b* & 14966 & 14.5& 79.8  & $ 5.99\pm 0.66$ & $ 0.26\pm 0.13$ & $ 5.73\pm 0.64$ & $0.91^{+0.05}_{-0.04}$ & $1.13^{+1.64}_{-1.50}$ & $6.4^{+6.3}_{-4.5}$ \\
\hline
J1544+0446a  & 14968 & 14.9& 5.8   & $ 0.38\pm 0.16$ & $ 0.32\pm 0.14$ & $<$ 0.58          & $-0.62^{+0.09}_{-0.38}$ & - & -  \\
J1544+0446b* & 14968 & 14.9& 49.9  & $ 3.93\pm 0.52$ & $ 0.45\pm 0.17$ & $ 3.48\pm 0.49$ & $0.76^{+0.10}_{-0.08}$ & $1.66^{+1.47}_{-1.35}$ & $3.9^{+3.2}_{-2.7}$ \\
\enddata
\tablecomments{* represent the targets with sufficient net counts for spectral analysis (see Section \ref{subsec:xspec}). (2) \chandra\ observation ID; (3) \chandra\ effective exposure, in units of ks; (4) Observed net counts in 0.5-8 keV band; (5)-(7) Observed count rate in 0.5-8 ($F$), 0.5-2 ($S$) and 2-8 ($H$) keV bands, in units of $10^{-3}{\rm~counts~s^{-1}}$; (8) Hardness ratio, defined as $HR$ = $(H-S)/(H+S)$; (9) Best-fitted photon index of a power-law model; (10) Best-fitted intrinsic column density of a power-law model, in units of $10^{22}{\rm~cm^{-2}}$. 
}
\end{deluxetable*}

\section{Observations, Data Reduction, and Data Analysis}\label{sec:obs}

\subsection{\chandra\ ACIS X-ray Imaging Spectroscopy}\label{subsec:chandra}

We observed the five dual-AGN candidates with the ACIS-S on board the \chandra\ X-ray Observatory between 2012 December and 2013 April (Program GO-14700264, PI: X. Liu).  All targets were observed on-axis on the S3 chip, 5.2$''$ to 15.0$''$ away from the aimpoint. Each target was observed for 15 ks (Table \ref{table:xray}). The exposure time was set to obtain $\sim$50 counts in the 2--10 keV range from the \OIII\ weaker nucleus in each merger, although the estimate was too optimistic. We estimated the X-ray counts from the \OIII\ luminosity  for each nucleus (corrected for \OIII\ emission due to star formation in AGN-H {\tiny II} composites \citep{kauffmann09}), assuming an empirical correlation between the 2--10 keV (unabsorbed) and \OIII\ luminosities. Measurements of $L_{\rm 2-10\,keV}/L_{{\rm [O\,{\tiny III}]}}$ for optically selected Type 2 AGNs span a wide range \citep{mulchaey94,heckman05,panessa06}, with values from a few to a few hundred. For the baseline assumption, we adopted the mean calibration of \citet{panessa06} given by, 
\begin{equation}
\log \bigg[\frac{L_{\rm 2-10\,keV}}{{\rm erg}\,{\rm s}^{-1}}\bigg] = (1.22\pm0.06)\log \bigg[\frac{L_{{\rm [O\,{\tiny III}]}}}{{\rm erg}\,{\rm s}^{-1}}\bigg] + (- 7.34\pm2.53), 
\end{equation}
where the \OIII\ luminosities have been corrected for the Galactic and intrinsic NLR extinction by using the Balmer decrement method via \halpha/\hbeta~ ratio. We accounted for systematic uncertainties using the \citet{heckman05} relation based on optically selected (single) Type 2 AGNs. An X-ray power-law spectrum was assumed with an absorbing column density $N_{\rm H}=10^{22}\ {\rm cm^{-2}}$ \citep[typical for our targets for which enough counts were detected for spectral analysis (see below) and for Type 2 Seyferts;][]{bassani99} and a photon index $\Gamma=1.7$ \citep[typical for unabsorbed Seyferts;][]{green09}. 


We reprocessed the data using CIAO v4.8 and the corresponding calibration files following standard procedure\footnote{http://cxc.harvard.edu/ciao/}. We examined the light curve of each observation and found no time interval of high background. We produced counts and exposure maps with the original pixel scale ($0 \farcs 492 {\rm~pixel^{-1}}$) in the 0.5--2 keV ($S$), 2--8 keV ($H$), and 0.5--8 keV ($F$) bands. The exposure maps were weighted by the above fiducial incident spectrum.


Following the source detection procedure detailed in \cite{Wang2004} and \citet{Hou2017}, we detect X-ray sources in the $S$, $H$ and $F$ bands in each image. With a local false detection probability $P \leq {10^{-6}}$ (empirically yielding $\sim$0.1 false detection per field), we detected a total of 124 sources in the field-of-view covered by the S3 and S2 CCDs. For each detected source, we derived background-subtracted and exposure map-corrected count rates in each individual band from within the 90\% enclosed-energy radius (EER), taking into account the position-dependent point-spread function and the local background. 

Figure \ref{fig:sdssimg} shows the ACIS images of the five targets in the $F$, $S$ and $H$ bands. Table \ref{table:xray} summarizes the X-ray measurements. Given the low count levels, we do not apply any smoothing to avoid artifacts. As we will show in Section \ref{subsec:xraylumi}, our targets are significantly weaker hard X-ray emitters than those predicted from both the \citet{panessa06} and \citet{heckman05} relations based on single optically selected AGNs, resulting in far fewer counts than expected. Seven of the ten nuclei in our targets were detected in the $F$-band. Six nuclei were detected in both $S$ and $H$ bands, whereas one nucleus (J1544+0446a) was detected in the $S$ band only. Three nuclei (J0805+2818b, J1330$-$0036a and J1058+3144a) were undetected in the X-rays.

\subsection{Spectral Analysis and Hardness Ratio}\label{subsec:xspec}

\begin{figure}
 \centering
 \includegraphics[width=0.42\textwidth, angle=0]{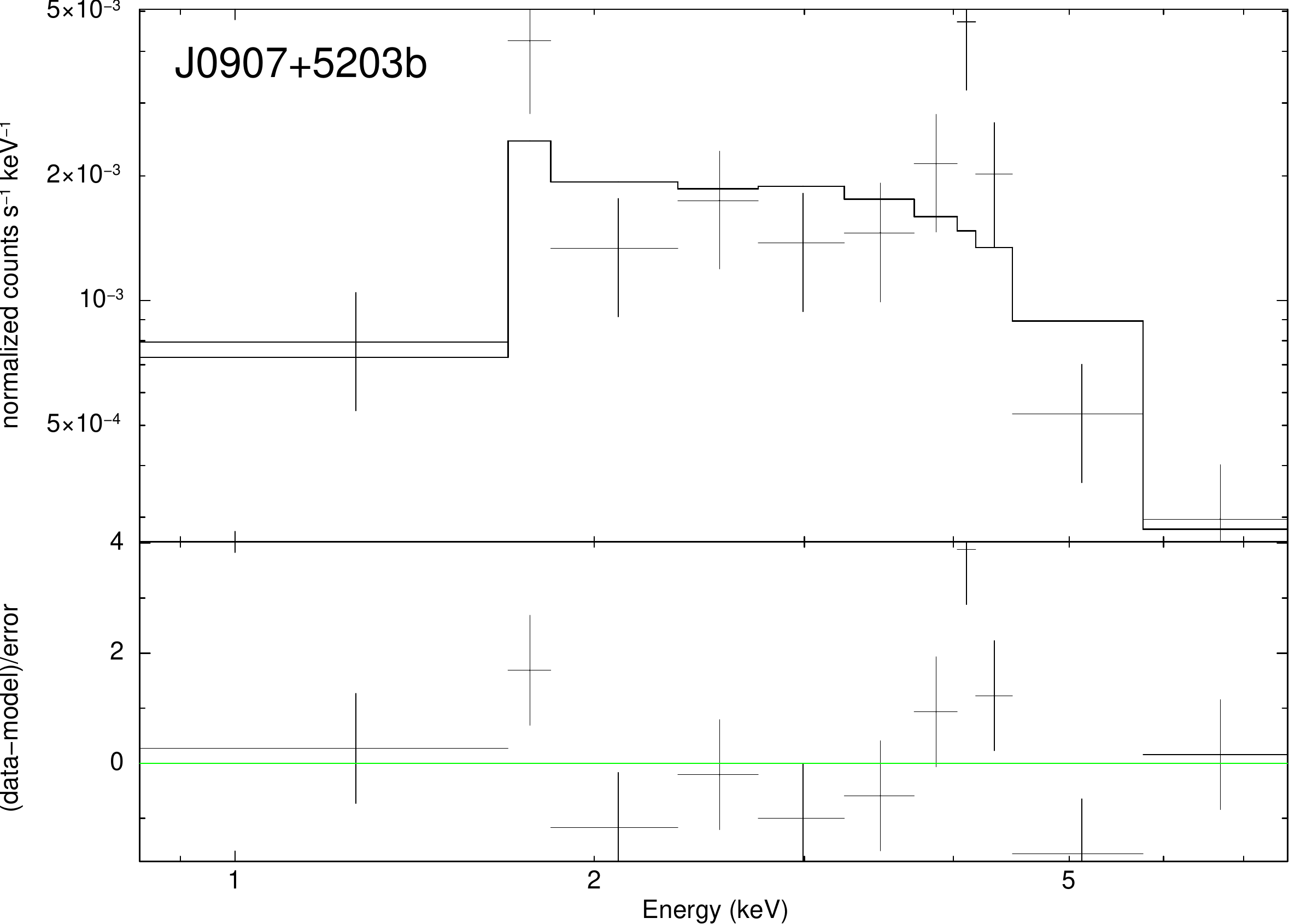}
 \includegraphics[width=0.42\textwidth, angle=0]{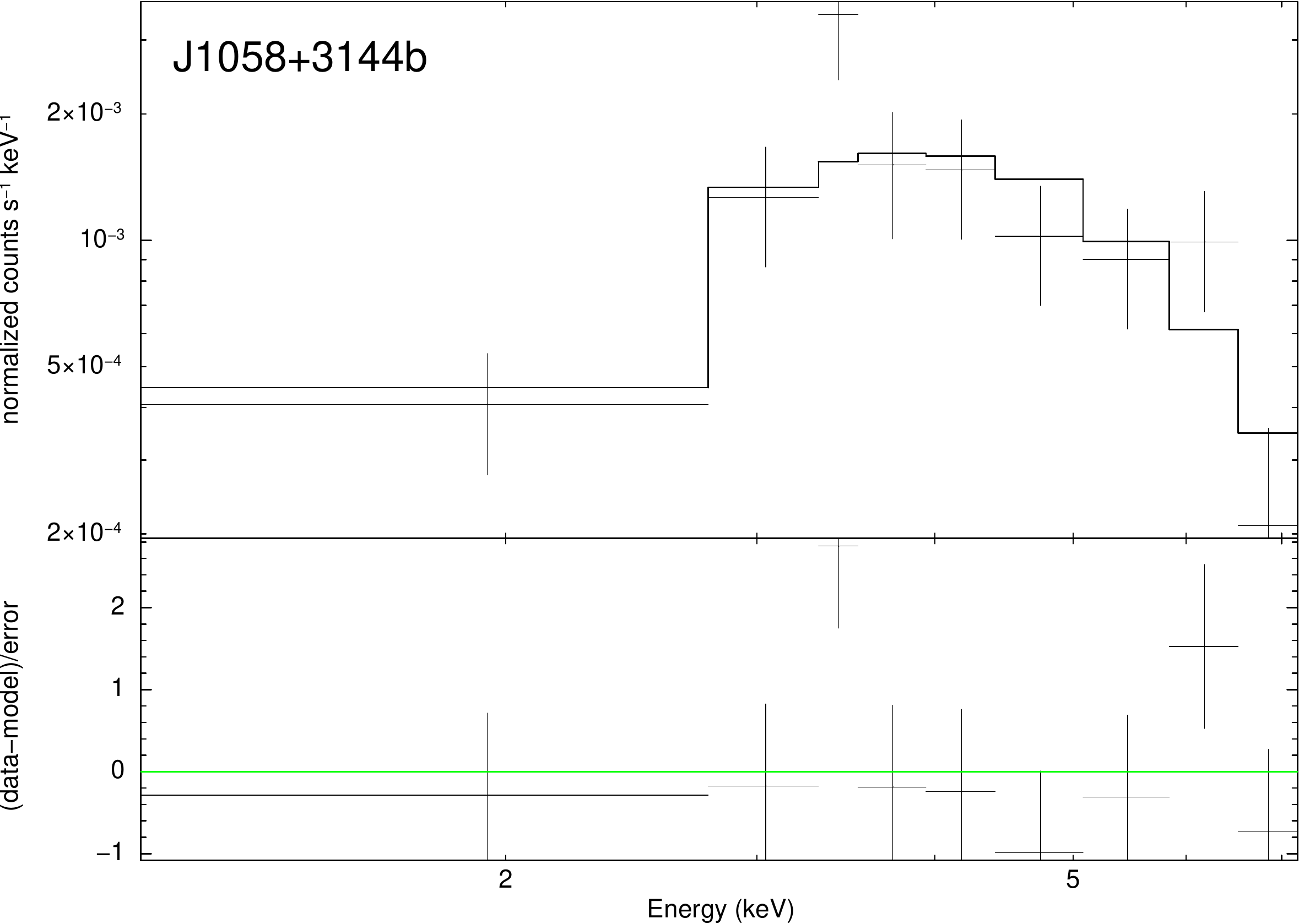}
 \includegraphics[width=0.42\textwidth, angle=0]{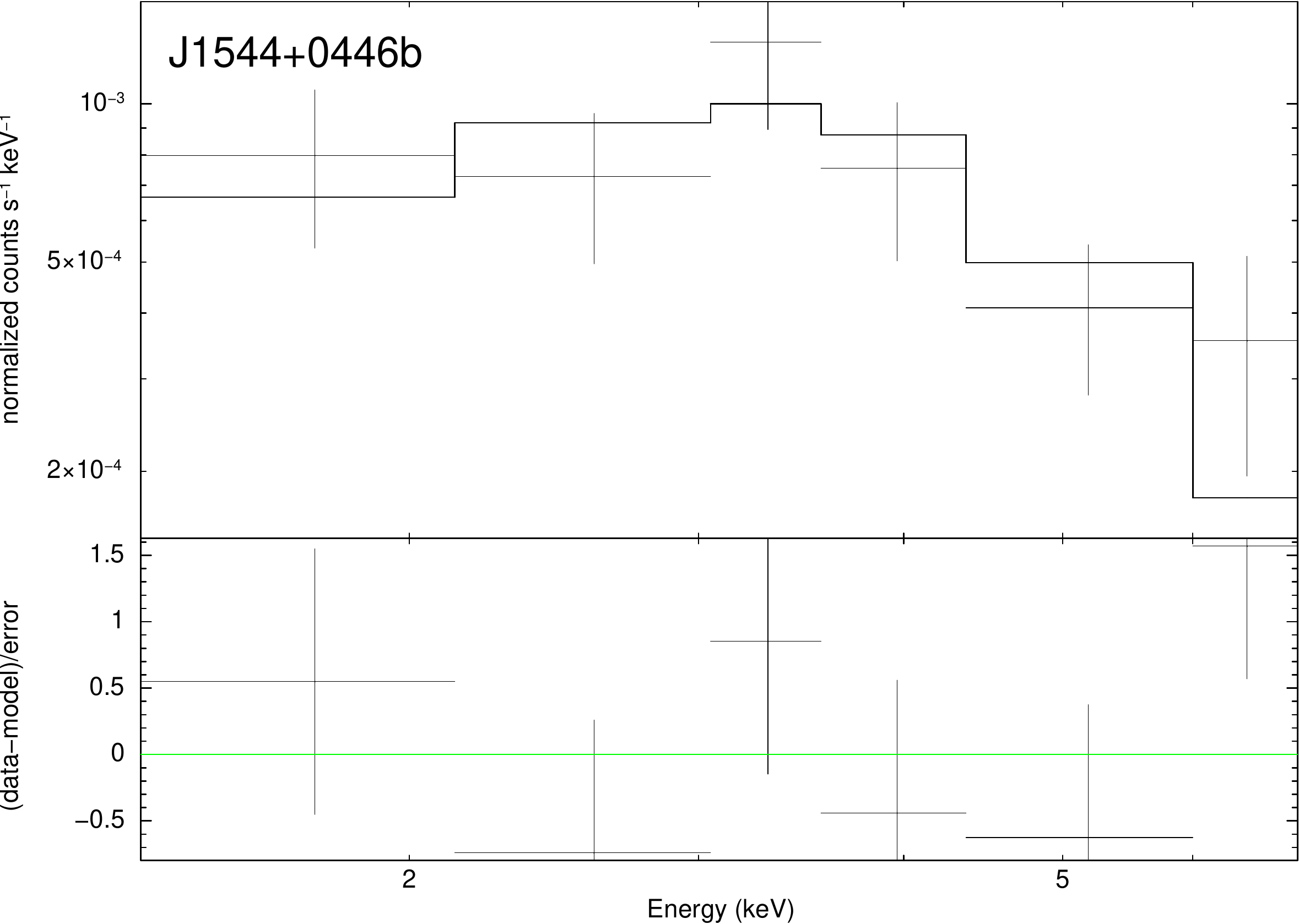}
\caption{Spectra modeling for three X-ray detected nuclei with $>50$ net counts. The spectra have been grouped to achieved a signal-to-noise ratio of 3. The fitted model is an absorbed power law. The best-fit photon index and intrinsic column density are listed in Table \ref{table:xray}. For each nucleus the lower panel shows the sigma of residuals.
    }\label{fig:spectra}
\end{figure}

Three of the seven X-ray detected nuclei have sufficient net counts (${\geq}50$) for spectral analysis. For these nuclei, we extracted spectra from within the 90\% EER of each source of interest using the CIAO tool {\it specextract} and built the Response Matrix Files (RMFs) and Ancillary Response Files (ARFs); the corresponding background spectrum was extracted typically between two to five times the 90\% EER. For nucleus J1544+0446b, with the other X-ray detected nucleus located in the background extraction region given the small projected nuclear separation (4.1\arcsec), we further removed the 90\% EER of this other nucleus from the background to eliminate its contamination. We fit each spectrum with an absorbed power-law model. Due to the low net counts, we adopted the C-statistic (a Poisson log-likelihood function, taking into account the known Poisson background, pstat in XSpec) for the spectral fitting \citep{cash79}. Figure \ref{fig:spectra} shows the fitted spectra. Table \ref{table:xray} lists the spectral fitting results. The best-fit power-law spectral index and the absorption column density are consistent with Type-2 AGN for all three nuclei.

For the other four X-ray detected AGNs with $<50$ net counts, we estimate their spectral properties using hardness ratios (HR). The hardness ratio is defined as
\begin{equation}
{\rm HR} \equiv \frac{H-S}{H+S},
\end{equation}
where $H$ and $S$ are the number of counts in the hard and soft X-ray bands, respectively. We adopt the Bayesian Estimation of Hardness Ratios \citep[BEHR;][]{park06} to measure the HRs and their uncertainties, which is appropriate for the low count regime. Table \ref{table:xray} lists the HR results.

For the three sources without X-ray detection and one nucleus (J1544+0446a) without X-ray detection in H band, we estimate the X-ray net counts 3$\sigma$ upper limit using the CIAO tool {\it aprate}.



\section{Results}\label{sec:result}

\subsection{X-ray Luminosities}\label{subsec:xraylumi}

X-ray emission provides the most direct evidence for nuclear activity. The 2--10 keV hard X-ray band is transparent to column densities of $N_{{\rm H}}\lesssim10^{24}$ cm$^{-2}$. To infer X-ray luminosity, we assume a simple absorbed power-law model. For the three nuclei with enough counts for X-ray spectral analysis, we adopt the best-fit column density and photon index given in Table \ref{table:xray}. For the other seven nuclei without enough counts or without X-ray detection, we assume a photon index of 1.7 and an absorption column density $N_{{\rm H}}=10^{22}$ cm$^{-2}$.
In Table \ref{table:xrayflux} we list the 
unabsorbed/intrinsic X-ray luminosity or upper limit of each nucleus in the total, soft, and hard bands, respectively. 

The adopted single absorbed power-law model is most likely too simple for the X-ray spectra of obscured AGNs, in which thermal emission from starburst components and/or scattered nuclear emission are often present \citep[e.g.,][]{turner97,turner97b}. However, the low counts of our detections do not allow us to test more realistic models. In addition, our estimates of the intrinsic absorbing column may not necessarily reflect the true values in cases of patchy obscuration and/or significant scattering off of an ionized medium in Compton-thick (i.e., $N_{{\rm H}}\sim10^{24}$ cm$^{-2}$ or larger) AGNs [which represent about half of the local Type 2 Seyfert population \citep{risaliti99}], as observed in NGC 6240 \citep[e.g.,][]{vignati99,ptak03} and in NGC 1068 \citep[e.g.,][]{matt97,guainazzi99}, although, again, the quality of our data do not allow us to robustly test these possibilities.

\begin{figure*}
 \centering
    \includegraphics[width=0.95\textwidth]{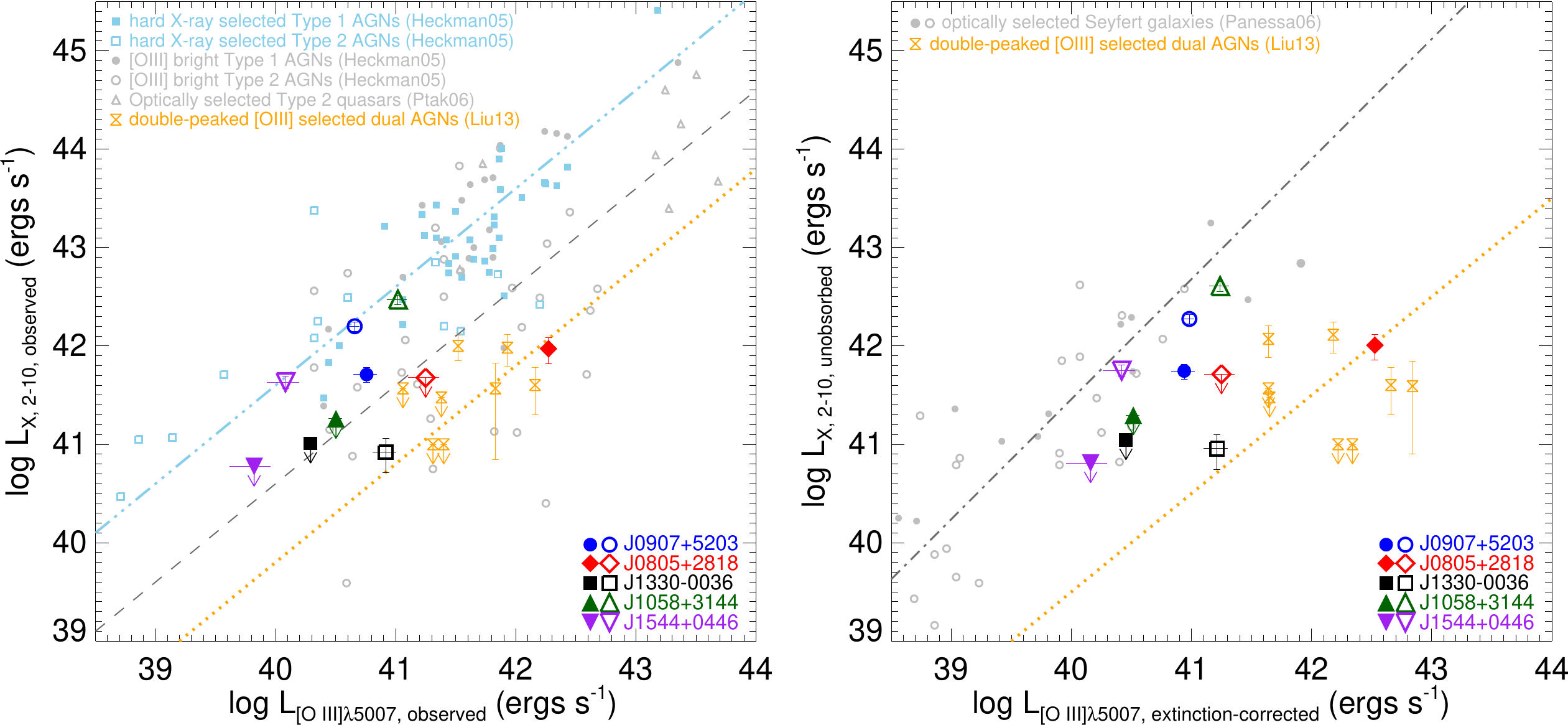}
    \caption{Hard X-ray luminosities vs. \OIII\ luminosities. Left panel: observed 2-10 keV luminosity vs. observed \OIII\ luminosity. For comparison, hard X-ray selected AGNs, \OIII\ bright AGNs \citep{heckman05}, optically selected Type 2 quasars \citep{ptak06} and double-peaked \OIII-selected dual AGN sample of \citet{Liu2013} are shown in sky blue squares (Type 1s as filled and Type 2s as open), grey circles (Type 1s as filled and Type 2s as open), grey open upward triangles and yellow hourglasses, respectively. 
    The mean relation for hard X-ray selected AGNs (both Type 1 and Type 2) and optically selected Type 1 AGNs \citep{heckman05}, optically selected Type 2 AGNs \citep{heckman05} and double-peaked \OIII-selected dual AGNs \citep{Liu2013} are shown in blue dashed-dotted-dotted, grey dashed and yellow dotted lines.
    Right panel: unabsorbed 2-10 keV luminosity vs. extinction-corrected \OIII\ luminosity. For comparison, nearby optically selected Seyfert galaxies \citep{panessa06} are shown in grey circles.
    The grey dashed-dotted line is the mean relation for mixed Seyferts in nearby galaxies \citep{panessa06} and the yellow dotted line is that for double-peaked \OIII\-selected dual AGNs \citep{Liu2013}.
    }\label{fig:lxlo3ratio}
\end{figure*}

The seven X-ray detected nuclei have estimated unabsorbed 0.5--8 keV luminosities ranging from $3.1 \times 10^{40}{\rm~erg~s^{-1}}$ to $4.1 \times 10^{42}{\rm~erg~s^{-1}}$. One nucleus J1544+0446a is not detected in the hard band, and the remaining six hard X-ray detected nuclei have estimated unabsorbed 2--10 keV luminosities ranging from $9.0 \times 10^{40}{\rm~erg~s^{-1}}$ to $4.0 \times 10^{42}{\rm~erg~s^{-1}}$. Figure \ref{fig:lxlo3ratio} shows the comparison of \OIII\ luminosities (both observed and extinction-corrected) and 2--10 keV X-ray luminosities (both observed and unabsorbed). The estimated upper limits for the four hard X-ray undetected nuclei range from $6.5 \times 10^{40}{\rm~erg~s^{-1}}$ to $5.1 \times 10^{41}{\rm~erg~s^{-1}}$ in 2--10 keV. These luminosity estimates are comparable to or smaller than those of the previously known X-ray confirmed dual AGNs \citep[e.g.,][]{komossa03,ballo04,hudson06,bianchi08,mazzarella11,Koss2011}.

\begin{deluxetable*}{lccccccccc}
\tablecaption{X-ray Luminosity of the Five \OIII -Selected Dual AGN Targets. \label{table:xrayflux}}
\tabletypesize{\footnotesize}
\tablehead{
\colhead{Name} & 
\colhead{log${L_{X,0.5-8}}$} & 
\colhead{log${L_{X,0.5-2}}$} & 
\colhead{log${L_{X,2-10}}$} & 
\colhead{log${L_{X,2-10,\rm obs}}$} & 
\colhead{log${L_{\rm \OIII, obs}}$} &  
\colhead{log${L_{\rm \OIII, cor}}$} & 
\colhead{SFR} & 
\colhead{log${L^{\rm SF}_{\rm 0.5-2}}$} & 
\colhead{log${L^{\rm SF}_{\rm 2-10}}$} 
}
\colnumbers
\startdata
J0907+5203a&	 $41.67^{+ 0.08}_{- 0.06}$ & $40.68^{+ 0.26}_{- 0.16}$ & $41.75^{+ 0.08}_{- 0.07}$ & $41.71^{+ 0.08}_{- 0.07}$ & $40.76^{+ 0.09}_{- 0.11}$ & $40.94^{+ 0.09}_{- 0.11}$ & $ 0.59^{+ 1.61}_{- 0.52}$ & $39.42^{+ 0.57}_{- 0.94}$ & $39.47^{+ 0.57}_{- 0.94 }$ \\
J0907+5203b*&	 $42.35^{+ 0.04}_{- 0.04}$ & $41.89^{+ 0.10}_{- 0.08}$ & $42.27^{+ 0.05}_{- 0.04}$ & $42.20^{+ 0.05}_{- 0.04}$ & $40.66^{+ 0.04}_{- 0.05}$ & $40.98^{+ 0.04}_{- 0.05}$ & $ 1.28^{+ 2.95}_{- 0.94}$ & $39.76^{+ 0.52}_{- 0.58}$ & $39.81^{+ 0.52}_{- 0.58 }$ \\
	\hline
J0805+2818a&	 $42.24^{+ 0.09}_{- 0.07}$ & $41.92^{+ 0.12}_{- 0.09}$ & $42.01^{+ 0.15}_{- 0.11}$ & $41.97^{+ 0.15}_{- 0.11}$ & $42.27^{+ 0.03}_{- 0.03}$ & $42.53^{+ 0.03}_{- 0.03}$ & $11.68^{+15.90}_{- 6.95}$ & $40.72^{+ 0.37}_{- 0.39}$ & $40.77^{+ 0.37}_{- 0.39 }$ \\
J0805+2818b&	 $< 41.87$                 & $< 41.81$                 & $< 41.71$                 & $< 41.68$                 & $41.25^{+ 0.11}_{- 0.15}$ & $41.25^{+ 0.11}_{- 0.15}$ & $ 2.24^{+ 3.31}_{- 1.38}$ & $40.00^{+ 0.39}_{- 0.41}$ & $40.05^{+ 0.39}_{- 0.41 }$ \\
	\hline   
J1330$-$0036a&  $< 40.90$                 & $< 40.71$                 & $< 41.05$                 & $< 41.01$                 & $40.29^{+ 0.02}_{- 0.02}$ & $40.45^{+ 0.02}_{- 0.02}$ & $ 0.35^{+ 0.49}_{- 0.21}$ & $39.20^{+ 0.38}_{- 0.39}$ & $39.24^{+ 0.38}_{- 0.39 }$ \\
J1330$-$0036b&  $41.16^{+ 0.13}_{- 0.10}$ & $40.84^{+ 0.18}_{- 0.13}$ & $40.96^{+ 0.21}_{- 0.14}$ & $40.92^{+ 0.21}_{- 0.14}$ & $40.92^{+ 0.09}_{- 0.11}$ & $41.21^{+ 0.09}_{- 0.11}$ & $ 3.98^{+ 6.69}_{- 2.57}$ & $40.25^{+ 0.43}_{- 0.45}$ & $40.30^{+ 0.43}_{- 0.45 }$ \\
	\hline
J1058+3144a&	 $< 41.32$                 & $< 41.20$                 & $< 41.30$                 & $< 41.26$                 & $40.50^{+ 0.05}_{- 0.06}$ & $40.51^{+ 0.05}_{- 0.06}$ & $ 1.95^{+ 2.73}_{- 1.19}$ & $39.94^{+ 0.38}_{- 0.41}$ & $39.99^{+ 0.38}_{- 0.41 }$ \\
J1058+3144b*&	 $42.61^{+ 0.05}_{- 0.05}$ & $42.12^{+ 0.31}_{- 0.18}$ & $42.61^{+ 0.05}_{- 0.05}$ & $42.47^{+ 0.05}_{- 0.05}$ & $41.02^{+ 0.08}_{- 0.09}$ & $41.24^{+ 0.08}_{- 0.09}$ & $ 2.44^{+ 5.33}_{- 1.75}$ & $40.04^{+ 0.50}_{- 0.55}$ & $40.09^{+ 0.50}_{- 0.55 }$ \\
	\hline 
J1544+0446a&	 $40.49^{+ 0.24}_{- 0.15}$ & $40.35^{+ 0.26}_{- 0.16}$ &  $<40.81$                 &  $<40.77$                 & $39.82^{+ 0.14}_{- 0.20}$ & $40.16^{+ 0.14}_{- 0.20}$ & $ 0.37^{+ 1.09}_{- 0.34}$ & $39.22^{+ 0.59}_{- 1.07}$ & $39.27^{+ 0.59}_{- 1.07 }$ \\
J1544+0446b*&	 $41.87^{+ 0.06}_{- 0.05}$ & $41.43^{+ 0.21}_{- 0.14}$ & $41.75^{+ 0.07}_{- 0.06}$ & $41.63^{+ 0.07}_{- 0.06}$ & $40.08^{+ 0.11}_{- 0.15}$ & $40.42^{+ 0.11}_{- 0.15}$ & $ 0.91^{+ 2.27}_{- 0.72}$ & $39.61^{+ 0.54}_{- 0.68}$ & $39.66^{+ 0.54}_{- 0.68 }$ \\
\enddata
\tablecomments{(2)-(4) Unabsorbed luminosity in 0.5-8 ($F$), 0.5-2 ($S$) and 2-10 keV bands. The luminosity of targets with * are derived from the fitted spectrum, while the others are converted by assuming an absorbed power-law with a photon index of 1.7 and an absorption column density ${N_{\rm H}} = 10^{22} {\rm~cm^{-2}}$; (5) Observed luminosity in 2-10 keV bands; (6)-(7) Observed and extinction-corrected \OIII\ luminosity; (8) Fiber star formation rate in units of $\rm{M}_{\odot}{\rm~{yr}^{-1}}$ given by the MPA-JHU DR7 catalog inferred from D$_n$(4000); (9)-(10) 0.5-2 ($S$) and 2-10 keV bands X-ray luminosities due to star formation. 
}
\end{deluxetable*}


\subsection{X-ray Contribution from Nuclear Starburst}\label{subsec:sf}

The estimated intrinsic hard X-ray luminosities of our targets are close to or below $\sim10^{42}$ ergs s$^{-1}$ -- the characteristic upper limit for the most luminous star-forming galaxies \citep[e.g.,][]{zezas01}. Hence it is possible that much or all of the luminosity is due to star formation. X-ray spectral shape offers another diagnostic to discriminate between AGN and starburst scenarios. However, the uncertainties of our spectral estimates are too large to draw firm conclusions for the majority of the nuclei.

We use independent star formation rate (SFR) estimates to test the AGN scenario for each nucleus. To estimate the expected X-ray emission due to star formation within similar apertures used to perform our X-ray extraction (typically a radius of 2\arcsec), we use the SDSS fiber SFR given by the MPA-JHU DR7 catalog \citep{salim07}. 
For galaxies classified as AGNs or composites according to the optical BTP diagram, which are the case for our targets, the SFRs are estimated by constructing the likelihood distribution of the specific SFR as a function of the 4000 \angstrom\ break D$_n$(4000) based on the star-forming sample \citep{brinchmann04} multiplied by the stellar mass.

\begin{figure*}
 \centering
    \includegraphics[width=0.95\textwidth]{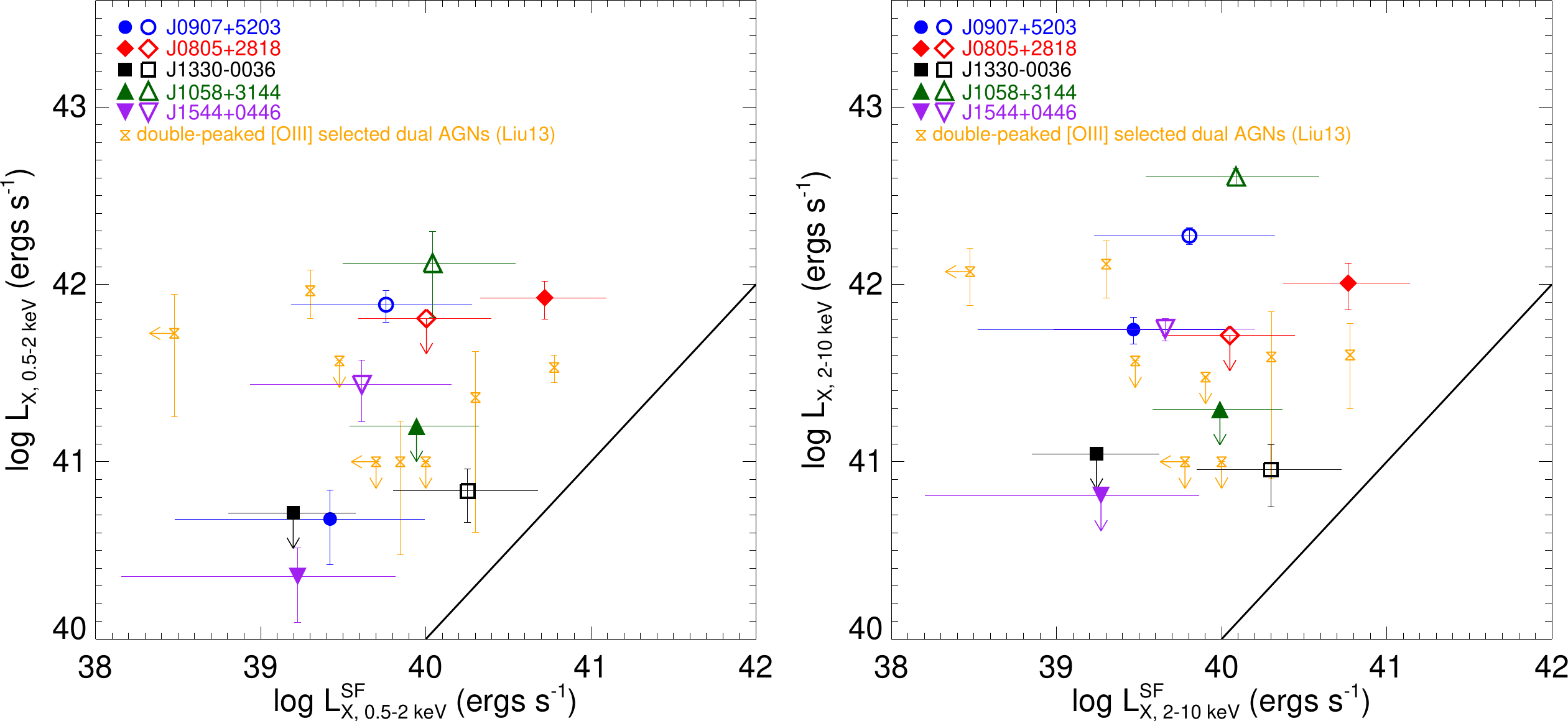}
    \caption{X-ray luminosities vs. the expected contribution from star-formation-related processes.  Left panel for soft band and right panel for hard band. The solid line shows the equality relation. Also shown for comparison (in yellow hourglasses) are the double-peaked \OIII -selected dual AGN sample of \citet{Liu2013}.
    }\label{fig:lxlxsf}
\end{figure*}

To derive X-ray luminosites from the fiber SFRs, we adopt the empirical calibration of \citet[][see also \citealt{grimm03}]{ranalli03} based on 23 nearby star-forming galaxies, which is given by
\begin{equation}\label{eq:lxs_sfr}
L^{{\rm SF}}_{0.5-2\, {\rm keV}} = 4.5 \times 10^{39} \frac{{\rm SFR}}{M_{\odot}~{\rm
yr}^{-1}} {\rm erg~s}^{-1}, \\
\end{equation}
\begin{equation}\label{eq:lxh_sfr}
L^{{\rm SF}}_{2-10\, {\rm keV}} = 5.0 \times 10^{39} \frac{{\rm SFR}}{M_{\odot}~{\rm
yr}^{-1}} {\rm erg~s}^{-1},
\end{equation}
with an rms scatter of 0.27 dex and 0.29 dex. In Table \ref{table:xrayflux} we list the fiber SFR estimates and the derived $L^{{\rm SF}}_{0.5-2\, {\rm keV}}$ and $L^{{\rm SF}}_{2-10\, {\rm keV}}$ estimates for each nucleus. Figure \ref{fig:lxlxsf} compares the expected X-ray luminosities due to star formation against the observed X-ray luminosities in the soft and hard bands. In both bands, the predicted X-ray contribution from star formation for the majority of our targets' nuclei is below the observed X-ray luminosity, suggesting an additional excitation source from the AGN. We caution, however, that there are significant systematic uncertainties of our estimates of the expected X-ray luminosities due to star-formation-related processes \citep[e.g., uncertainties in the IMF, extinction correction;][]{Liu2013}.

We also test the contribution to hard X-ray luminosity from low-mass X-ray binaries, which is proportional to stellar mass ($M_*$). Based on the $L_{\rm X}-M_{\ast}$ relations from \citet{Gilfanov2004} or \citet{Lehmer2010}, the estimated contribution from stellar mass enclosed in the SDSS fiber is negligible compared to the contribution from star forming activity (typically $<$ 10\% and never exceeding 30\%).

\subsection{Results on Individual Targets}\label{subsec:nature}



\subsubsection{SDSS J0907+5203}

Both galaxies in the merger are optically classified as Type 2 Seyferts (Figure \ref{fig:bpt}). Both nuclei were detected in both soft and hard X-ray bands. The northern galaxy (J0907+5203b) has enough counts for X-ray spectral analysis, which suggests moderate nuclear obscuration, with an estimated column density of $N_{{\rm H}}{\approx}2.6^{+1.5}_{-1.1}\times10^{22}$ cm$^{-2}$. The spatial profiles of the two nuclear X-ray sources are consistent with the AGN scenarios for both galaxies. For both galaxies, the expected star-formation-induced X-ray luminosities are too low to explain the observed values in both soft and hard X-ray bands (Figure \ref{fig:lxlxsf}), consistent with the dual AGN scenario.  

\subsubsection{SDSS J0805+2818}

The NW nucleus (J0805+2818a) in the merger is optically classified as a Type 2 Seyfert whereas the SE nucleus (J0805+2818b) a Type 2 Seyfert or a LINER (Figure \ref{fig:bpt}). Only the NW nucleus was detected in both soft and hard X-ray bands; the SE nucleus was detected in neither. The host galaxy of the SE nucleus shows strong Balmer absorption features in the SDSS spectrum characteristic of post-starburst galaxies (Figure \ref{fig:specfit1}). For the NW nucleus, both the spatial profile of the X-ray source and the comparison between the observed to the expected star-formation-induced X-ray luminosities (Figure \ref{fig:lxlxsf}) are consistent with the AGN scenario. For the SE nucleus, on the other hand, the upper limits of the X-ray luminosities (Figure \ref{fig:lxlxsf}) are still consistent with the presence of an additional AGN, although the possibility that one AGN in the NW nucleus ionizes gas in both galaxies cannot be ruled out.

\subsubsection{SDSS J1330$-$0036}

The NW nucleus (J1330$-$0036a) in the merger is optically classified as an H {\tiny II}/AGN composite whereas the SE nucleus (J1330$-$0036b) a Type 2 Seyfert (Figure \ref{fig:bpt}). Only the SE nucleus was detected in both soft and hard X-ray bands; the NW nucleus was not detected in either. The host galaxy of the NW nucleus shows strong Balmer absorption features in the SDSS spectrum characteristic of post-starburst galaxies (Figure \ref{fig:specfit1}). For the SE nucleus, the observed X-ray luminosities are similar to those expected from star-formation-induced X-ray luminosities (Figure \ref{fig:lxlxsf}), although an additional AGN component cannot be ruled out given significant uncertainties in the estimates. For the NW nucleus, on the other hand, the upper limits of the X-ray luminosities (Figure \ref{fig:lxlxsf}) are still consistent with the presence of an additional AGN, although the possibility that one AGN in the SE nucleus ionizes gas in both galaxies cannot be ruled out.

\subsubsection{SDSS J1058+3144}

The SW nucleus (J1058+3144a) in the merger is optically classified as an H {\tiny II}/AGN composite whereas the NE nucleus (J1058+3144b) a Type 2 Seyfert (Figure \ref{fig:bpt}). Only the NE nucleus was detected in both soft and hard X-ray bands; the SW nucleus was not detected in either. The NE nucleus has enough counts for X-ray spectral analysis, which suggests moderate nuclear obscuration, with an estimated column density of $N_{{\rm H}}{\approx}6.4^{+6.3}_{-4.5}\times10^{22}$ cm$^{-2}$. For the NE nucleus, both the spatial profile of the X-ray source and the comparison between the observed to the expected star-formation-induced X-ray luminosities (Figure \ref{fig:lxlxsf}) are consistent with the AGN scenario. For the SW nucleus, on the other hand, the upper limits of the X-ray luminosities (Figure \ref{fig:lxlxsf}) are still consistent with the presence of an additional AGN, although the possibility that one AGN in the NE nucleus ionizes gas in both galaxies cannot be ruled out.

\subsubsection{SDSS J1544+0446}

The SW nucleus (J1544+0446a) in the merger is optically classified as a LINER whereas the NE nucleus (J1544+0446b) a LINER, Type 2 Seyfert or composite (Figure \ref{fig:bpt}). The NE nucleus was detected in both soft and hard X-ray bands; the SW nucleus was detected only in the soft X-ray band. The NE nucleus has enough counts for X-ray spectral analysis, which suggests moderate nuclear obscuration, with an estimated column density of $N_{{\rm H}}{\approx}3.9^{+3.2}_{-2.7}\times10^{22}$ cm$^{-2}$. For the NE nucleus, both the spatial profile of the X-ray source and the comparison between the observed to the expected star-formation-induced X-ray luminosities (Figure \ref{fig:lxlxsf}) are consistent with the AGN scenario. An additional AGN component is likely also for the SW nucleus because the observed soft X-ray luminosity is significantly larger than that expected from star-formation-related activity, despite significant uncertainties.

\vspace{0.3cm}
It is noteworthy that those target galaxies without X-ray detection tend to have a blue color in the composite SDSS image (Figure \ref{fig:sdssimg}). These seem to be relatively small galaxies with young stellar populations. One possibility is that the X-ray emission from these galaxies is dominated by star formation and they are more likely to host a dwarf SMBH which is harder to detect. Whether this is a definitive trend can only be answered with a sizable sample of close pairs.

In summary, the new \chandra\ X-ray observations support the dual-AGN scenario for two of our five \OIII -selected targets (SDSS J0907+5203 and SDSS J1544+0446). For the other three targets (SDSS J0805+2818, SDSS J1330$-$0036, and SDSS J1058+3144), the existing data are still consistent with the dual-AGN scenario, although the possibility of only one AGN ionizing both components in the mergers cannot be ruled out.

\section{Discussion}\label{sec:discuss}

\subsection{Systematically Smaller X-ray-to-\OIII -luminosity Ratio in Dual AGNs than in Single AGNs}\label{subsec:x2oratio}

Figure \ref{fig:lxlo3ratio} shows the relation between the hard X-ray luminosity and the \OIII\ luminosity for each nucleus in our targets. We compare the X-ray-to-\OIII -luminosity ratio of our new targets studied in this work to those observed in both single AGNs and dual AGNs systematically selected from double-peaked \OIII\ emission lines \citet[][see also \citealt{wang09,Smith2010,Ge2012,LyuLiu2016,Yuan2016}]{Liu2010a}. We examine both the relation between the observed hard X-ray luminosity $L_{X,2-10\,{\rm keV, observed}}$ and the observed \OIII\ luminosity $L_{{\rm [O\,III],observed}}$, and that between the unabsorbed hard X-ray luminosity $L_{X,2-10\,{\rm keV, unabsorbed}}$ and the extinction-corrected \OIII\ luminosity $L_{{\rm [O\,III],extinction-corrected}}$. We use the appropriate comparison samples for the two cases separately, because usually either the observed or the corrected luminosity is available in any given literature sample.

For the $L_{X,2-10\,{\rm keV, observed}}$-$L_{{\rm [O\,III],observed}}$ relation (left panel of Figure \ref{fig:lxlo3ratio}), the comparison samples include 47 hard X-ray (3--20 keV) selected local AGNs and 55 optically selected local \OIII -bright AGNs \citep{xu99,whittle92} studied by \citet{heckman05}, and 8 optically selected Type 2 quasars from \citet{ptak06} at redshifts $z{\sim}$0.3--0.8. \citet{heckman05} showed that single, optically selected Type 2 AGNs (the grey dashed line) have systematically lower $L_{X,2-10\,{\rm keV, observed}}$ (by an average of 1.0 dex) at a given $L_{{\rm [O\,III],observed}}$ than hard X-ray selected AGNs (both Type 1 and Type 2) and optically selected Type 1 AGNs (the blue dashed-dotted-dotted line), as expected for heavily absorbed AGNs \citep[see also][]{mulchaey94,panessa06}. \citet{Liu2013} has shown that the four dual AGNs (individually as the yellow standing glasses and collectively as the yellow dotted line) selected from the parent sample of Type 2 AGNs with double-peaked \OIII\ emission lines \citep{Liu2010b} have systematically smaller $L_{X}/L_{{\rm [O\,III]}}$ (observed) ratios (by $\sim0.8\pm0.2$ dex on average) than even optically selected single Type 2 AGNs. Our new targets seem to have X-ray-to-\OIII -luminosity ratios that are on average in between that observed in single, optically selected Type 2 AGNs and that observed in the dual AGNs selected from double-peaked \OIII\ emission lines (with some ambiguities and uncertainties due to the upper limits of several measurements). 



A similar trend is also seen in the $L_{X,2-10\,{\rm keV, unabsorbed}}$-$L_{{\rm [O\,III],extinction-corrected}}$ relation as shown in the right panel of Figure \ref{fig:lxlo3ratio}). Again, our new targets seem to have X-ray-to-\OIII -luminosity ratios that are on average in between that observed in single AGNs (the grey dashed-dotted line) and that observed in dual AGNs selected from double-peaked \OIII\ emission lines (the yellow dotted line). The comparison sample of single AGNs includes 47 Palomar Seyfert galaxies (optically selected Type 1 and Type 2 Seyferts drawn from the Palomar survey of nearby galaxies by \citealt{ho95}) from \citet{panessa06}. \citet{panessa06} has demonstrated that after properly accounting for absorption correction (including for Compton-thick sources), optically selected Type 1 and Type 2 Seyferts follow the same $L_{X,2-10\,{\rm keV, unabsorbed}}$-$L_{{\rm [O\,III],extinction-corrected}}$ relation. In particular, optically selected Type 2 Seyferts, which were significantly X-ray weaker than Type 1 Seyferts, also obey the same relation, after the ``Compton thick'' luminosity correction. \citet{Liu2013} has shown that after correction for gas absorption and dust extinction, the unabsorbed hard X-ray luminosities of double-peaked-\OIII -selected dual AGNs appear to be $\sim2.4\pm0.3$ dex smaller (at log$L_{{\rm [O\,III]}}$ of 42.0) than those expected from the \citet{panessa06} relation, log$L_{X}$ = $1.22$log$L_{{\rm [O\,III]}}-7.34$, although the absorption correction of dual AGNs may have been significantly underestimated. 

\subsection{Interpretation: Enhanced Nuclear Absorption from Merger-induced Gas Inflows}

\citet{Liu2013} has suggested that the observed X-ray weak tendency in dual AGNs selected in Type 2 AGNs with double-peaked narrow \OIII\ lines is caused by a combination of a higher nuclear gas column, which may be induced by merger events, and an orientation bias related to the double-peak narrow emission-line selection. In contrast to the \citet{Liu2013} sample, our targets are not subject to the orientation bias due to the line-of-sight velocity splitting requirement caused by the double-peaked \OIII\ selection. On the other hand, our sample is likely to have a higher nuclear absorption from merger-induced gas inflows than that in single local AGNs, similar to the case of double-peaked-\OIII -selected dual AGNs.  Arising from the narrow-line regions that are much further out, the \OIII\ emission is less subject to nuclear gas absorption and dust obscuration than the hard X-ray emission from the black hole accretion disk corona, which would explain the systematically smaller hard-X-ray-to-\OIII -luminosity ratios observed in dual than in single AGNs. The fact that our targets seem to have hard-X-ray-to-\OIII -luminosity ratios that are smaller than that seen in single AGNs but larger than that observed in double-peaked-\OIII -selected dual AGNs is consistent with the conclusion of \citet{Liu2013} that a combination of two effects (i.e., both merger-enhanced absorption and obscuration and an orientation selection bias) are at work for double-peaked-\OIII -selected dual AGNs.


\section{Conclusions}\label{sec:sum}

Dual AGNs are crucial to our understanding of the accretion and dynamical evolution of SMBHs in mergers, the effects of merger-induced activity on galaxy evolution, and the initial conditions of close binary SMBHs. Building on \chandra~'s previous success on its unique power in resolving dual AGNs, here we have studied the X-ray properties of a sample of five optically selected dual AGN candidates. Our targets were drawn from a sample of 1286 \OIII -selected AGN pairs (both Type 1 and Type 2 sources) systematically selected from the SDSS DR7. Each of the targets contains two nuclei separated by 3--9 kpc in projection, both of which are optically classified as Type 2 (obscured) AGNs based on diagnostic ratios of the optical narrow emission lines. While being systematically selected from the largest sample of dual AGN candidates, the optical classification was inconclusive. Furthermore, because the double nuclei are close (with physical projected separations of a few kpc), there may be only one AGN ionizing both galaxies, producing two optical emission-line nuclei. Arguments based on the spatial distribution of ionization parameters estimated from optical emission lines cannot conclusively discriminate between the single- and dual-AGN scenarios \citep[e.g.,][]{Liu2010b}. The new \chandra\ ACIS-S X-ray imaging presented here helps solve the problem by resolving and localizing the ionizing sources directly in the X-rays. The X-ray confirmation of a systematically selected sample also helps place the optically inferred dual-AGN frequency on a firmer ground \citep{Liu2011a}. 
Our main findings are summarized as the following:

\begin{itemize}

\item \chandra~'s superb spatial resolution and sensitivity in the X-rays allowed us to localize the ionizing sources and determine their X-ray properties. Seven of the ten nuclei were detected in the full 0.5--8 keV band. Six were detected in both soft (0.5--2 keV) and hard (2--8 keV) bands, whereas one nucleus was detected in the soft band only. Three nuclei were undetected in the X-rays (Table \ref{table:xray}). 

\item The hard X-rays directly probe the accretion disk corona of the accreting SMBHs, providing a more robust estimate of the intrinsic AGN luminosity than using \OIIIb\ luminosity as a surrogate. In three of the ten nuclei we observed enough counts to perform spectral fittings to constrain the X-ray spectral properties and absorption column densities. We fit each X-ray spectrum with an absorbed power-law model. The best-fit power-law spectral indices and the absorption column densities are consistent with them being Type 2 AGNs for all three nuclei. For the other four X-ray detected nuclei, we have estimated their spectral properties and luminosities using hardness ratios. For the three sources without X-ray detection, we estimate the X-ray net counts 3$\sigma$ upper limit using the CIAO tool {\it aprate} (Section \ref{subsec:xspec}). 

\item Combined with independent star formation rate estimates empirically calibrated based on the host-galaxy stellar continua, the new \chandra\ X-ray observations allowed us to evaluate the dual-AGN hypothesis for each target. We have confirmed two (SDSS J0907+5203 and SDSS J1544+0446) of the five targets as bona-fide dual AGNs. For the other three targets, the existing data are consistent with the dual-AGN scenario, but we cannot conclusively rule out the possibility of stellar and/or shock heating and/or one AGN ionizing both gaseous components in a merger (Section \ref{subsec:nature}).  

\item The average X-ray-to-\OIII\ luminosity ratio in our targets seems to be systematically smaller than that observed in single AGNs but is higher than that seen in dual AGNs selected from AGNs with double-peaked narrow emission lines. We suggest that the systematically smaller X-ray-to\OIII\ luminosity ratio observed in dual AGNs than in single AGNs is due to a high nuclear gas column likely from strong merger-induced inflows. Unlike double-peaked-\OIII -selected dual AGNs, the new sample selected from resolved galaxy pairs are not subject to the orientation bias caused by the double-peak line-of-sight velocity splitting selection, which also contributes to lowering the X-ray-to-\OIII\ luminosity ratio (Figure \ref{fig:lxlo3ratio}).

\end{itemize}

Our sample size is still too small for a statistical analysis to compare with theoretical predictions from simulations in a meaningful way \citep[e.g.,][]{Capelo2017,Blecha2018,Rosas-Guevara2019,Solanes2019}. To put the conclusions on a firm statistical ground, future wide-field, high-resolution, and high-sensitivity X-ray telescopes \citep[such as Lynx X-ray Surveyor;][]{TheLynxTeam2018} may detect thousands of dual AGNs, which will be needed to fully understand black hole growth in mergers and dual AGNs \citep[e.g.,][]{Burke-Spolaor2018,Koss2019}.  

\section*{Acknowledgements}

We thank M. Strauss for his generous support and guidance on the project. M.H. and Z.L. acknowledge support by the National Key Research and Development Program of China (2017YFA0402703). X.L. and H.G. acknowledge support by NASA through \chandra\ Award Number GO3-14103X issued by the \chandra\ X-ray Observatory Center, which is operated by the Smithsonian Astrophysical Observatory for and on behalf of NASA under contract NAS 8-03060. Y.S. acknowledges support from the Alfred P. Sloan Foundation and NSF grant 1715579. 

This research has made use of software provided by the \chandra\ X-ray Center in the application packages CIAO, ChIPS, and Sherpa.

Funding for the Sloan Digital Sky Survey IV has been provided by the Alfred P. Sloan Foundation, the U.S. Department of Energy Office of Science, and the Participating Institutions. SDSS-IV acknowledges support and resources from the Center for High-Performance Computing at the University of Utah. The SDSS website is www.sdss.org.

SDSS-IV is managed by the Astrophysical Research Consortium for the Participating Institutions of the SDSS Collaboration including the Brazilian Participation Group, the Carnegie Institution for Science, Carnegie Mellon University, the Chilean Participation Group, the French Participation Group, Harvard-Smithsonian Center for Astrophysics, Instituto de Astrof\'isica de Canarias, The Johns Hopkins University, Kavli Institute for the Physics and Mathematics of the Universe (IPMU) / University of Tokyo, Lawrence Berkeley National Laboratory, Leibniz Institut f\"ur Astrophysik Potsdam (AIP),  Max-Planck-Institut f\"ur Astronomie (MPIA Heidelberg), Max-Planck-Institut f\"ur Astrophysik (MPA Garching), Max-Planck-Institut f\"ur Extraterrestrische Physik (MPE), National Astronomical Observatories of China, New Mexico State University, New York University, University of Notre Dame, Observat\'ario Nacional / MCTI, The Ohio State University, Pennsylvania State University, Shanghai Astronomical Observatory, United Kingdom Participation Group,Universidad Nacional Aut\'onoma de M\'exico, University of Arizona, University of Colorado Boulder, University of Oxford, University of Portsmouth, University of Utah, University of Virginia, University of Washington, University of Wisconsin, Vanderbilt University, and Yale University.

Facilities: \chandra\ X-ray Observatory (ACIS), Sloan

\section*{Appendix}

\begin{figure*}
 \centering
 \includegraphics[width=0.48\textwidth]{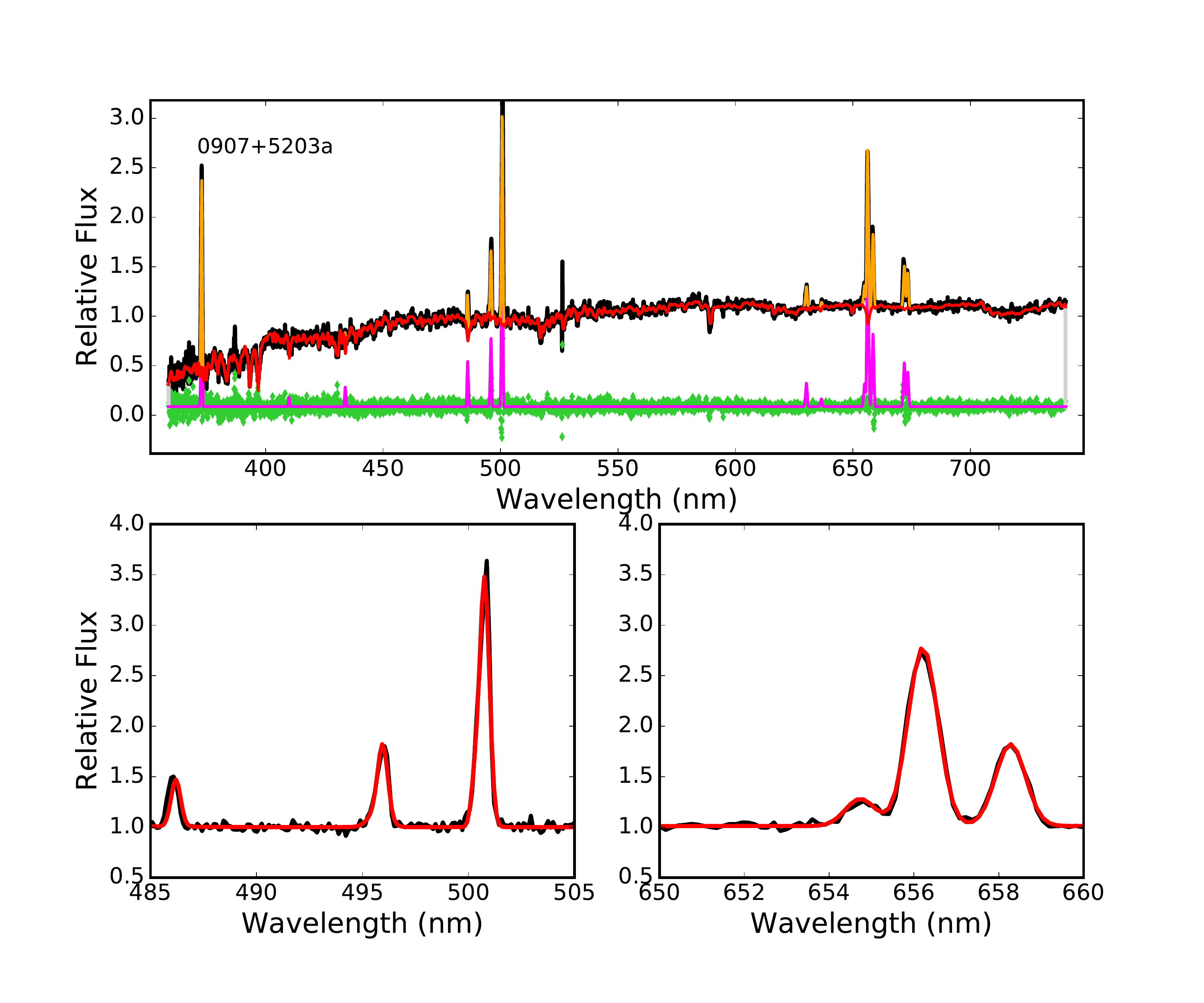}
 \includegraphics[width=0.48\textwidth]{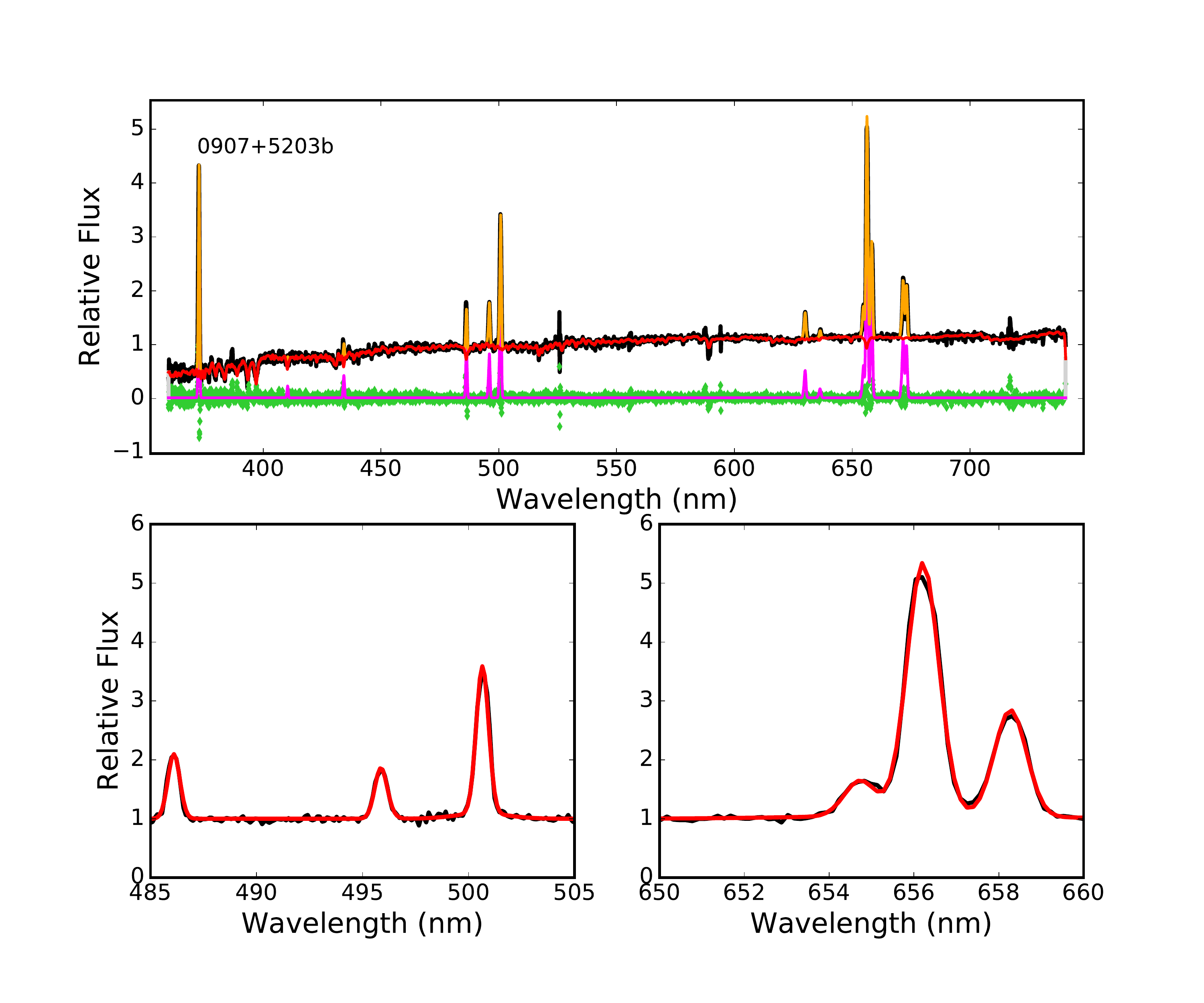}
 \includegraphics[width=0.48\textwidth]{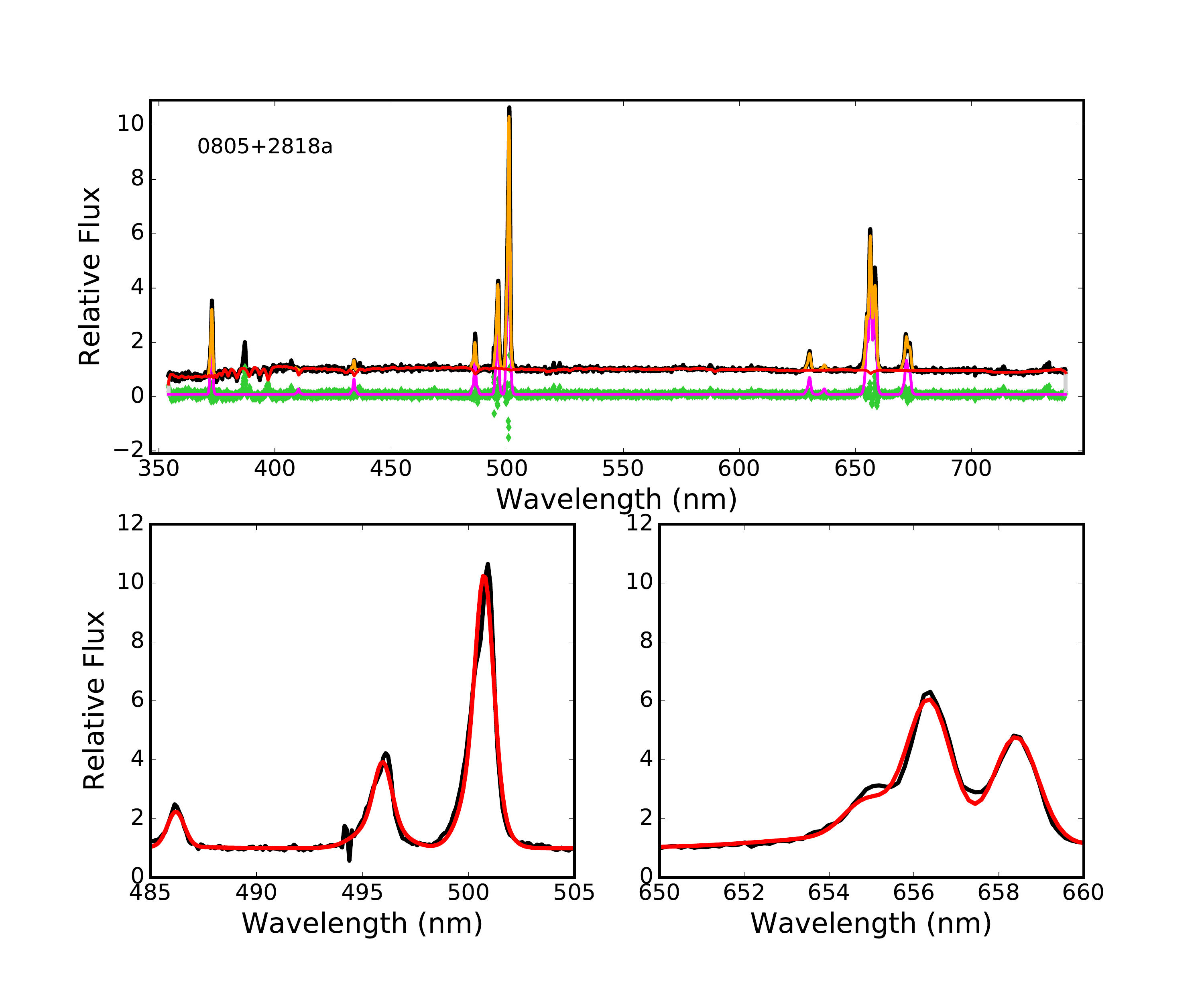}
 \includegraphics[width=0.48\textwidth]{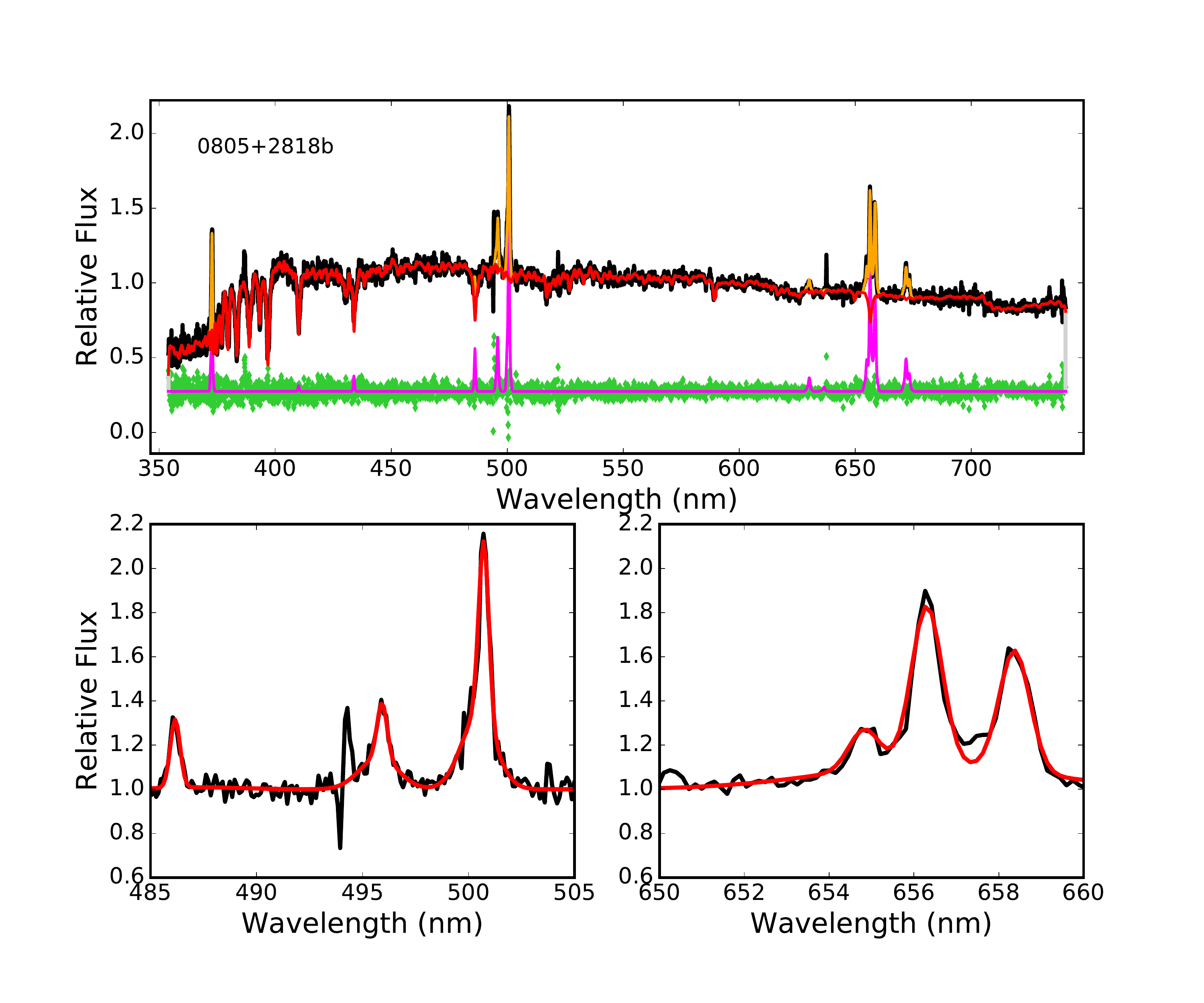}
 \includegraphics[width=0.48\textwidth]{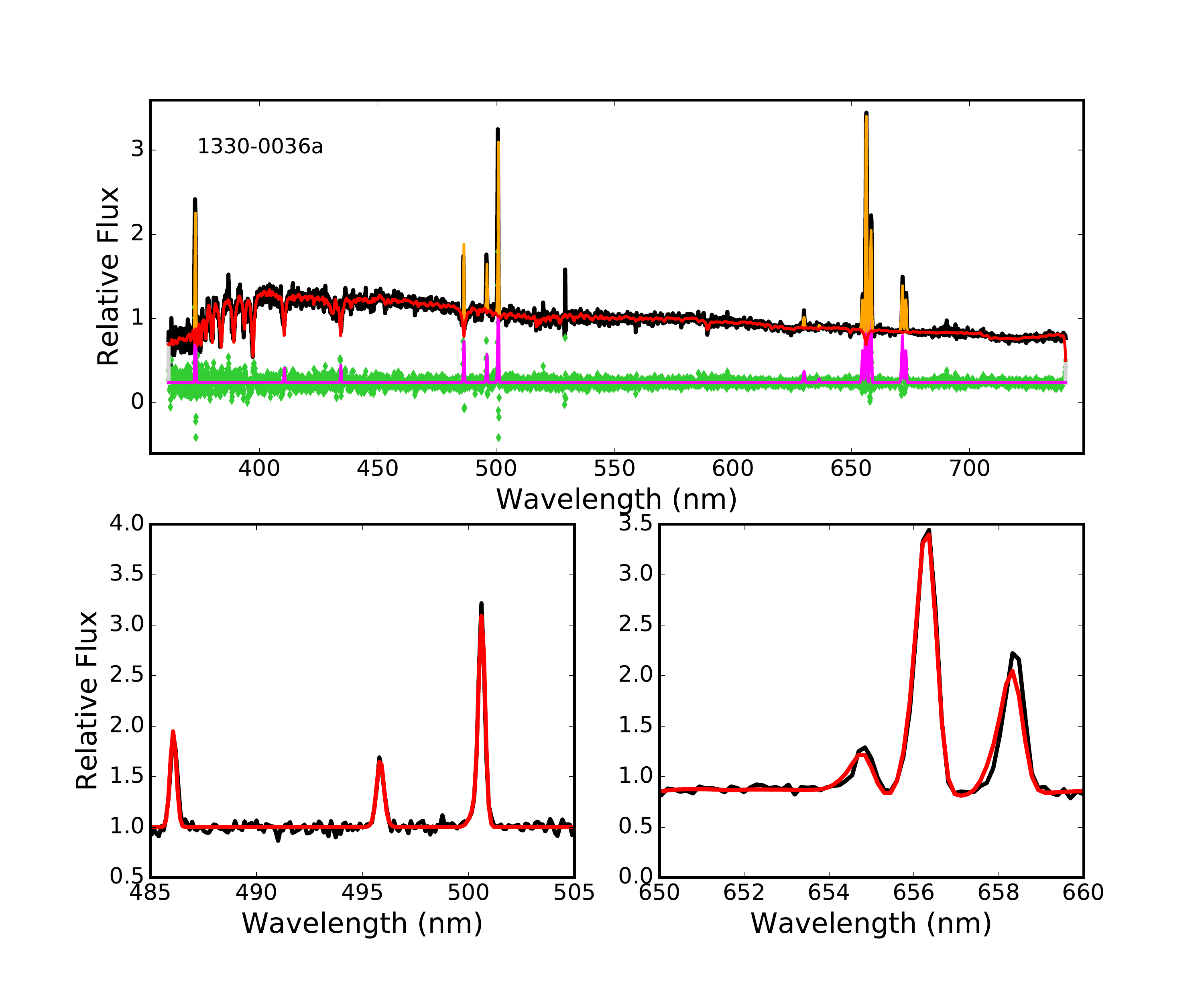}
 \includegraphics[width=0.48\textwidth]{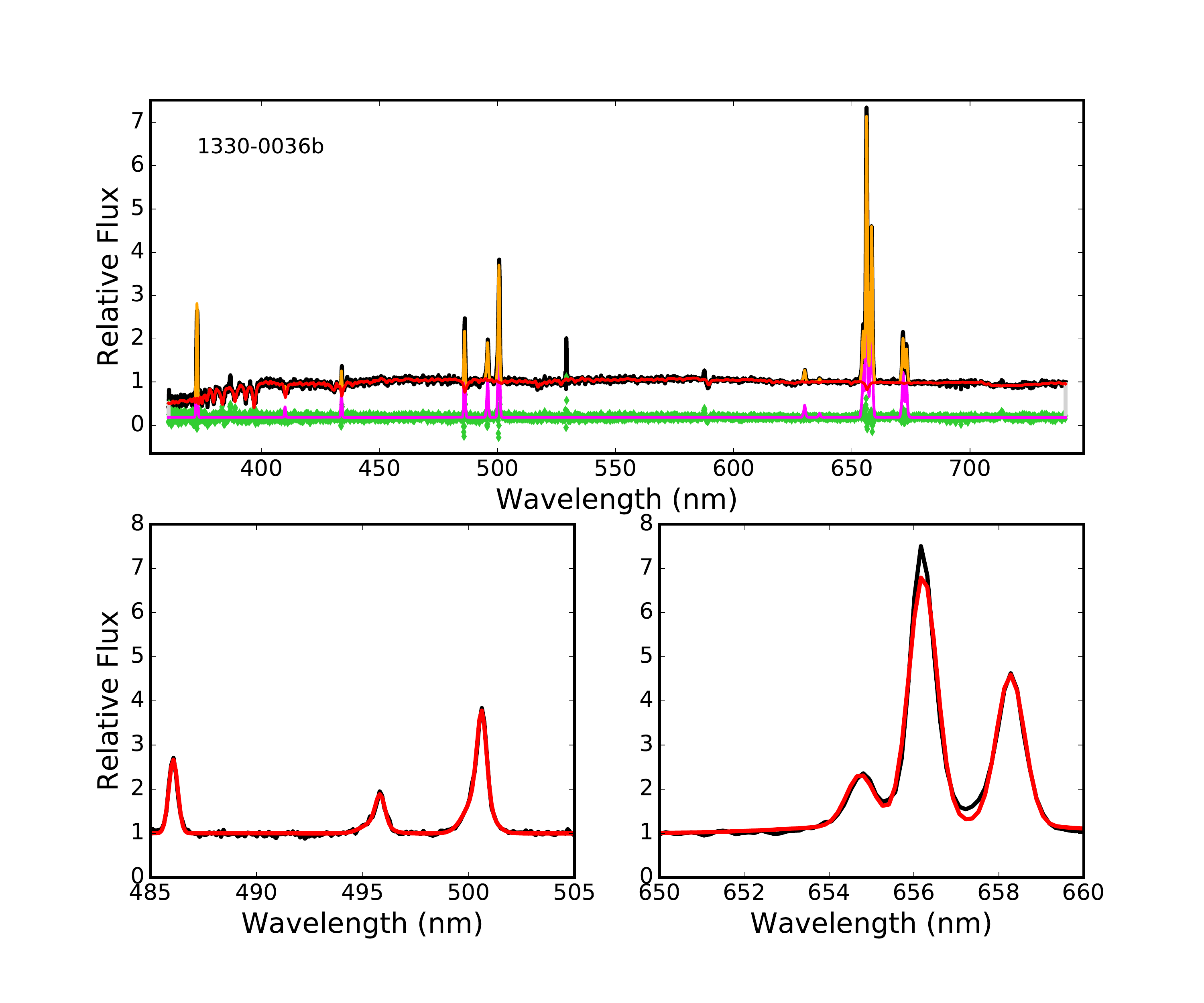}
    \caption{Spectral fitting results for the two nuclei in J0907+5203, J0805+2818 and J1330-0036 from top to bottom. For each nucleus, the upper panel shows the best-fit model from the pPXF fit for the host-galaxy stellar continuum shown in red overlaid on top of the SDSS spectrum shown in black. Also shown are the host-galaxy-subtracted emission-line spectrum in magenta, the total model (i.e., host+emission line) in orange, and the total residual (i.e., data-model) in green. The bottom panels show the best-fit model (in red) for the host-subtracted emission lines from the qsofit analysis overlaid on top of the data (in black).
    }\label{fig:specfit1}
\end{figure*}
\begin{figure*}
 \centering
 \includegraphics[width=0.48\textwidth]{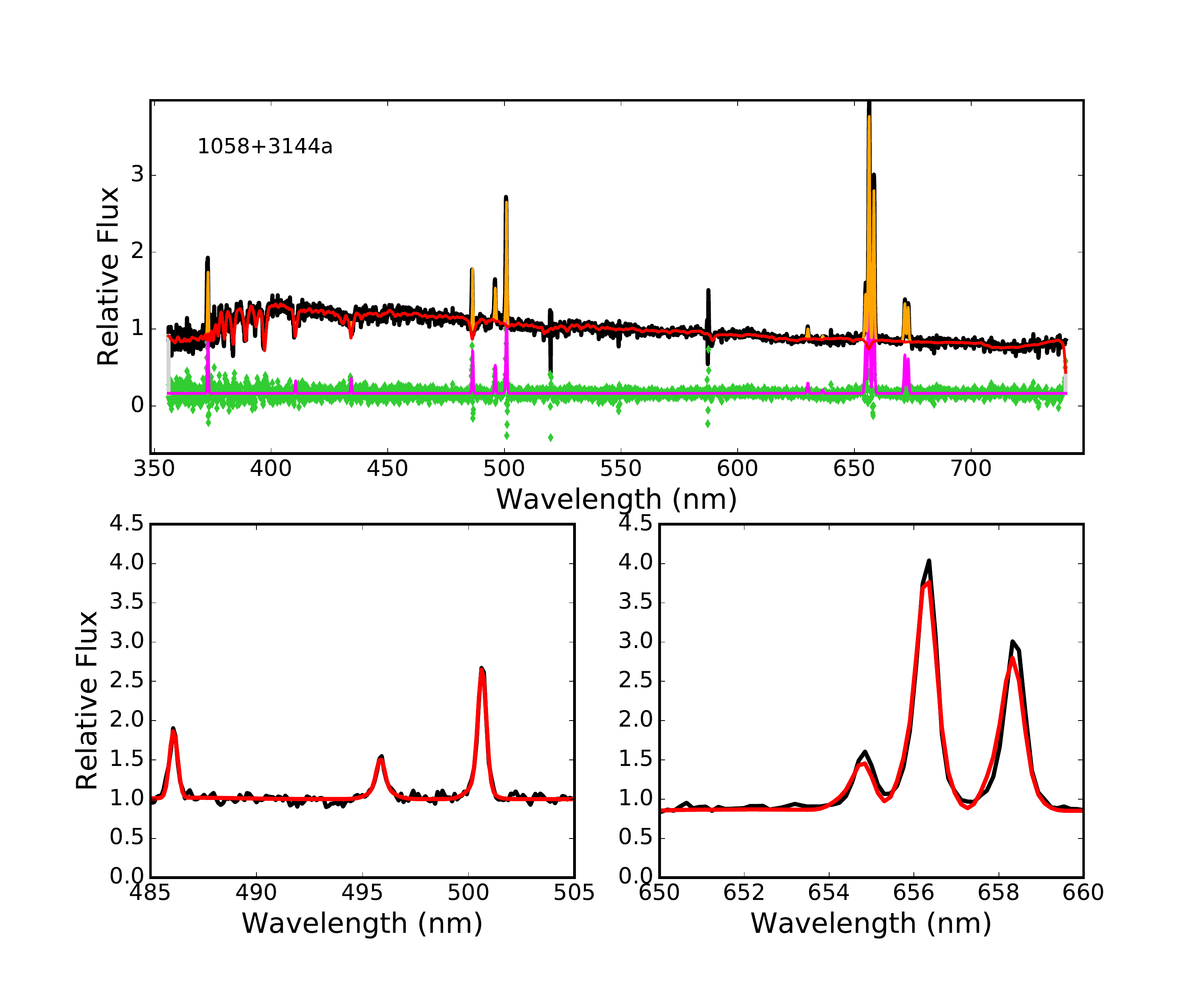}
 \includegraphics[width=0.48\textwidth]{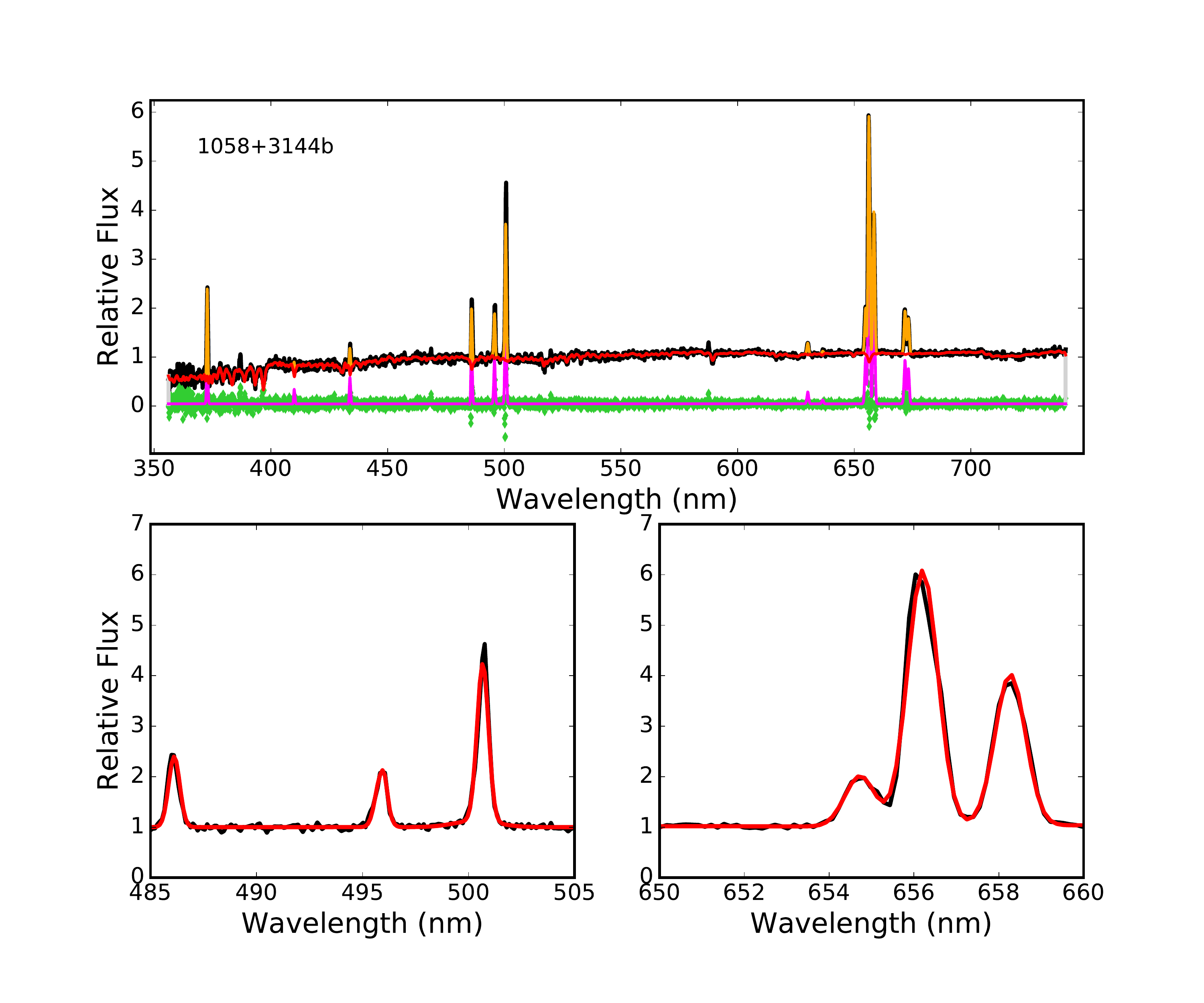}
 \includegraphics[width=0.48\textwidth]{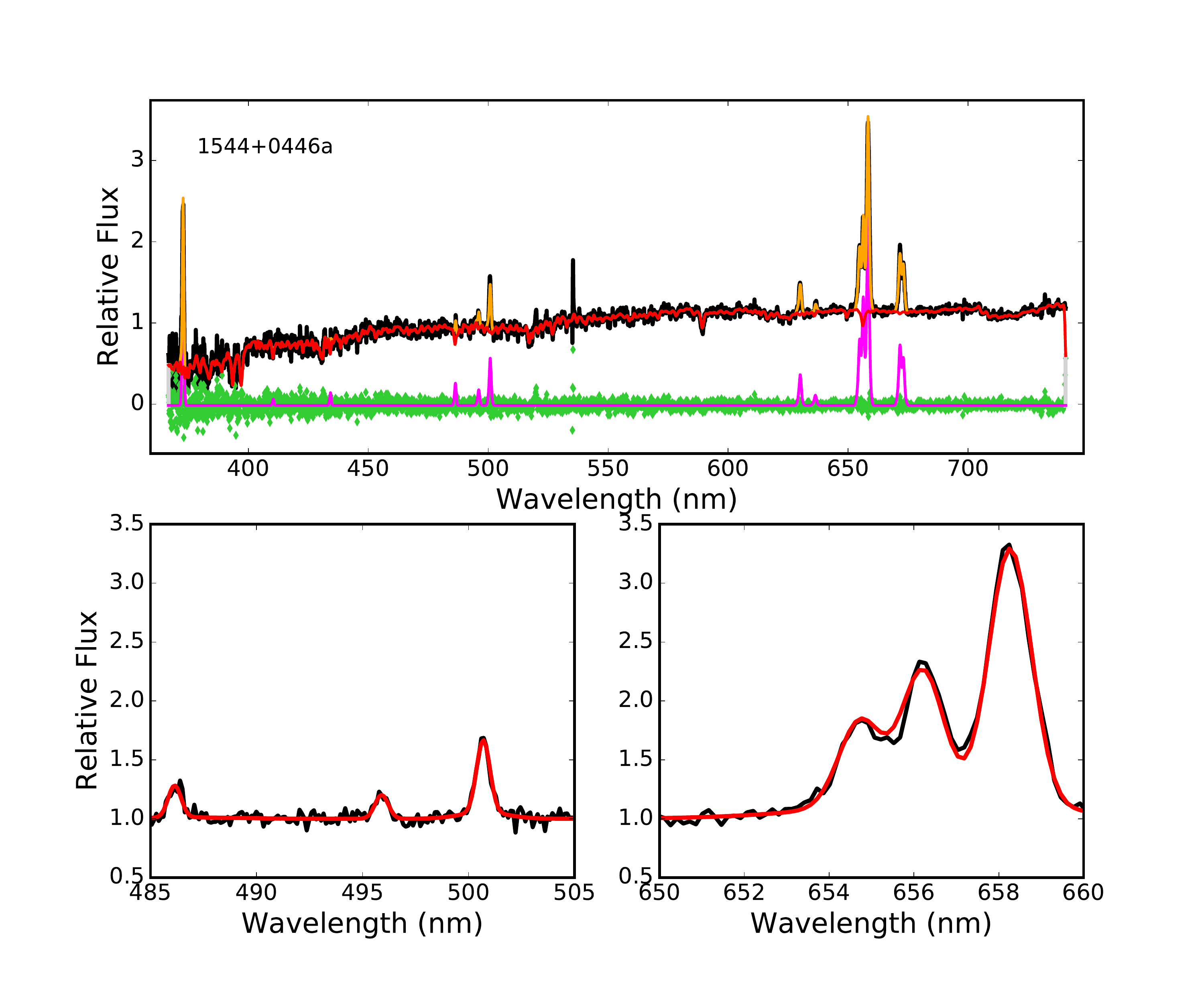}
 \includegraphics[width=0.48\textwidth]{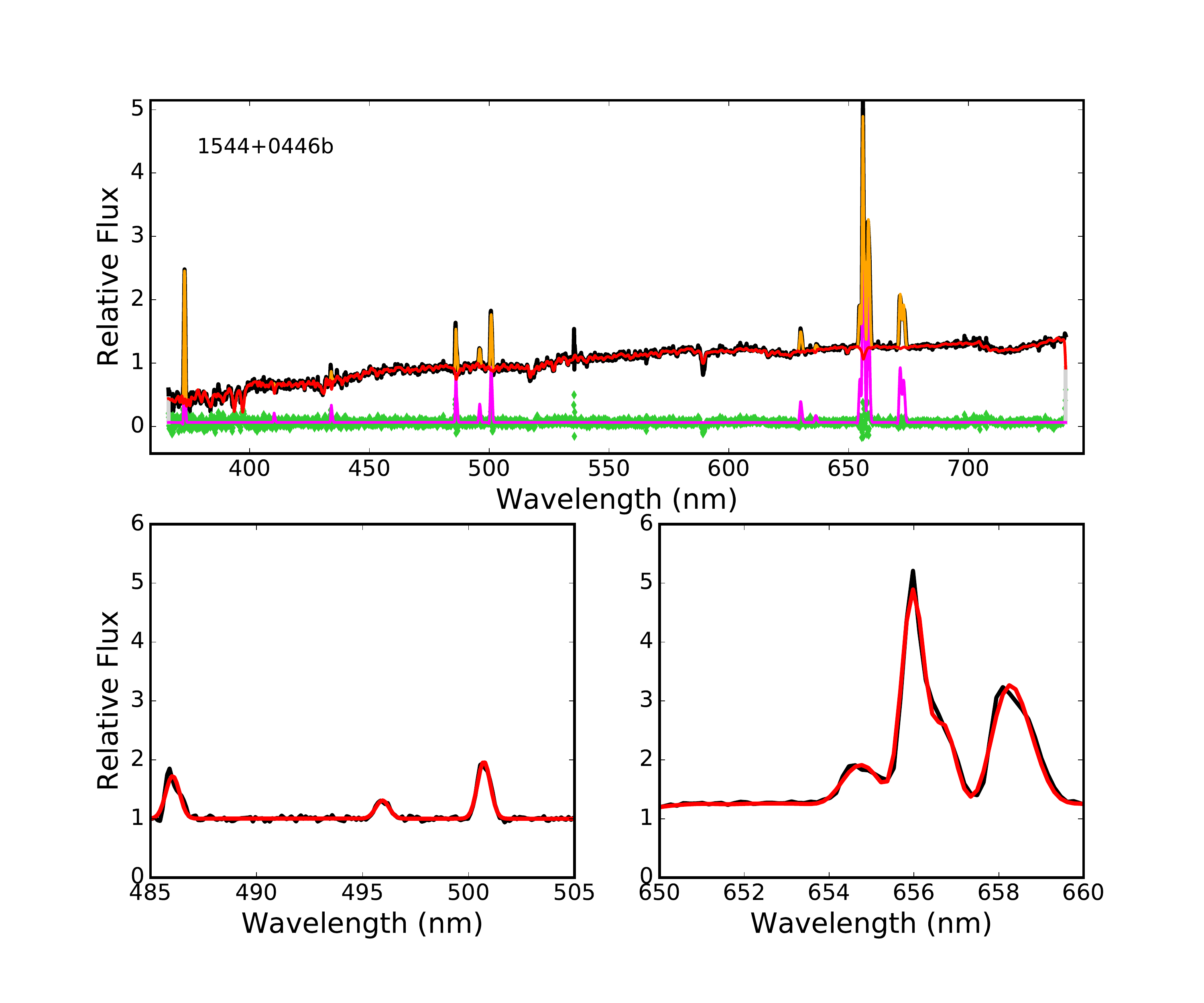}
    \caption{Same as Figure \ref{fig:specfit1}, but for J1058+3144 and J1544+0446. 
    }\label{fig:specfit2}
\end{figure*}

In this appendix, we present details of our spectral fitting analysis to carefully measure the host-galaxy stellar continuum and to model the emission-line fluxes over the host-subtracted spectrum. For the host galaxy spectral fitting, we adopt the penalized Pixel-Fitting (pPXF) method\footnote{https://pypi.org/project/ppxf/} \citep{Cappellari2004}. The method works directly in the pixel space and uses the maximum penalized likelihood formalism to extract as much information as possible from the spectra while suppressing the noise in the solution. After subtracting the host-galaxy continuum using the pPXF best-fit solution, we then model the emission-line-only spectrum using the spectral fitting code qsofit\footnote{https://github.com/legolason/PyQSOFit} \citep{Shen2018b}. Figures \ref{fig:specfit1} and \ref{fig:specfit2} show the fitting results for all the 10 nuclei in our targets.

\bibliography{binaryrefs}

\end{document}